\documentclass[11pt,fleqn]{article}

\usepackage[english]{babel}
\usepackage[utf8]{inputenc}
\usepackage{hyperref}

\usepackage{a4}
\usepackage{amstext}
\usepackage{amsfonts}
\usepackage{amssymb}
\usepackage{color, colortbl}
\usepackage{cite}
\usepackage{epsfig}
\usepackage{mathtools}
\usepackage{ulem}
\usepackage{caption}         
\usepackage{subcaption}   
\usepackage{framed}
\usepackage{multirow}
\usepackage{booktabs}

\usepackage{dsfont}
\usepackage{physics}
\usepackage{graphicx}
\usepackage[percent]{overpic}
\usepackage{rotating}
\usepackage{placeins}
\usepackage{authblk}

\newcommand{\RN}[1]{\uppercase\expandafter{\romannumeral#1}}

\newcommand{\gtapprox}{\raisebox{-0.5ex}{$\,\stackrel{>}{\scriptstyle\sim}\,$}}
\newcommand{\ltapprox}{\raisebox{-0.5ex}{$\,\stackrel{<}{\scriptstyle\sim}\,$}}

\usepackage{bm}

\def\j{{\cal J}}

\newcommand{\be}{\begin{equation}}
	\newcommand{\ee}{\end{equation}}
\newcommand{\bea}{\begin{eqnarray}}
	\newcommand{\eea}{\end{eqnarray}}

\usepackage[color=red!40, textsize=small]{todonotes}

\title{\bf Hybrid spin-dependent and hybrid-quarkonium mixing potentials at order $(1 /m_Q)^1$ from SU(3) lattice gauge theory}

\author[1]{Carolin Schlosser}
\author[1,2]{Marc Wagner}

\affil[1]{\normalsize Institut f\"ur Theoretische Physik, Goethe Universit\"at Frankfurt am Main, Max-von-Laue-Stra{\ss}e 1, D-60438~Frankfurt~am~Main, Germany}
\affil[2]{\normalsize Helmholtz Research Academy Hesse for FAIR, Campus Riedberg, Max-von-Laue-Stra{\ss}e 12, D-60438~Frankfurt~am~Main, Germany }

\date{January 15, 2025}

\begin{document}

	\maketitle
	
	\begin{abstract}
		We present the first lattice gauge theory results for hybrid spin-dependent and hybrid-quarkonium mixing potentials appearing at order $(1 /m_Q)^1$ in the Born-Oppenheimer Effective Field theory for hybrid mesons with gluon spin $\kappa^{PC} = 1^{+-}$.
		Specifically, we compute the four unknown potentials $V^{sa}_{11}(r)$, $V^{sb}_{10}(r)$, which are relevant for the hyperfine splitting in heavy hybrid meson spectra, as well as $V^\text{mix}_{\Sigma_u^-}(r)$ and $V^\text{mix}_{\Pi_u}(r)$, which describe the mixing of heavy hybrid mesons with ordinary quarkonium.
		We relate these potentials to matrix elements, which we extract from generalized Wilson loops with a chromomagnetic field insertion along one of the temporal lines and suitable hybrid creation operators replacing the spatial lines.
		We use gradient flow, which facilitates the renormalization of the matrix elements and has the additional benefit of significantly reducing statistical noise.
		We present results for gauge group SU(3) and lattice spacing $a=0.060 \, \text{fm}$.
		Our results demonstrate that a future combined continuum and zero-flow time extrapolation is possible within our setup, which will be necessary to reliably predict the hyperfine splitting of heavy hybrid mesons as well as their mixing with ordinary quarkonium through coupled channel Schrödinger equations.
	\end{abstract}

	
	\clearpage	

	\section{Introduction}
	Mesons are hadrons with integer spin, consisting typically of quark-antiquark pairs confined by gluons. If the glounic field is in an excited configuration and contributes to the quantum numbers of the system, one speaks of hybrid mesons.
	In this work we use lattice gauge theory to compute potentials at order $(1/m_Q)^1$ relevant for the study of heavy hybrid mesons, where the heavy quarks are typically $\bar b b$ or $\bar c c$ ($m_Q$ denotes the heavy quark mass, i.e.\ $m_Q = m_b$ or $m_Q = m_c$).

	The Born-Oppenheimer Effective Field Theory (BOEFT) provides a suitable framework to describe exotic hadrons such as heavy hybrid mesons, but also tetraquarks, doubly heavy baryons and pentaquarks (see e.g.\ Refs.\ \cite{Berwein:2015vca,Oncala:2017hop,Brambilla:2017uyf,Brambilla:2018pyn,Brambilla:2019jfi,Soto:2020xpm,Berwein:2024ztx}).
	Effective field theories like the BOEFT exploit the separation of scales, which are inherent in these systems, because the heavy quarks move non-relativistically within potentials depending on the relativistic light degrees of freedom, which can be either gluons or light quarks.
	The BOEFT Lagrangian is formulated as an expansion in terms of the inverse heavy quark mass $1/m_Q$ and potentials arise at each order in $1/m_Q$.
	The aim of this work is to carry out the first SU(3) lattice gauge theory computation of next-to-leading order (i.e.\ order $(1/m_Q)^1$) spin-dependent potentials and hybrid-quarkonium mixing potentials for heavy hybrid mesons with gluon spin $\kappa^{PC} = 1^{+-}$ (the lightest heavy hybrid mesons are expected to have $\kappa^{PC} = 1^{+-}$).
	
	The leading order (i.e.\ order $(1/m_Q)^0$) potentials relevant for heavy hybrid mesons are hybrid static potentials, which have been extensively studied using lattice gauge theory for both small and large quark-antiquark separations (see Refs.\ \cite{Griffiths:1983ah,
		Campbell:1984fe,
		Campbell:1987nv,
		Perantonis1989StaticPF,
		Michael:1990az,
		Perantonis:1990dy,
		Juge:1997nc,
		Peardon:1997jr,
		Juge:1997ir,
		Morningstar:1998xh,
		Michael:1998tr,
		Michael:1999ge,
		Juge:1999ie,
		Juge:1999aw,
		Bali:2000vr,
		Morningstar:2001nu,
		Juge:2002br,
		Juge:2003qd,
		Michael:2003ai,
		Michael:2003xg,
		Bali:2003jq,
		Juge:2003ge,
		Wolf:2014tta,
		Reisinger:2017btr,
		Bicudo:2018yhk,
		Bicudo:2018jbb,
		Reisinger:2018lne,
		Capitani:2018rox,
		Schlosser:2021hed,
		Schlosser:2021wnr,
		Sharifian:2023idc}).
	Hybrid static potentials, which are characterized by quantum numbers $\Lambda_\eta^\epsilon$, can be computed from the large time behavior of Wilson loop-like correlation functions, where suitable hybrid creation operators replace the straight spatial lines of ordinary Wilson loops.
	Lattice results for these hybrid static potentials have been used to predict spin-degenerate heavy hybrid meson spectra (see Refs.\ \cite{Perantonis:1990dy,Juge:1997nc,Juge:1999ie,Guo:2008yz,Braaten:2014qka,Berwein:2015vca,Oncala:2017hop,Capitani:2018rox,Brambilla:2018pyn,Brambilla:2019jfi,Schlosser:2021wnr}).
	These works utilized the leading order static potentials in either coupled channel Schrödinger equations or in combination with the simplifying single-channel approximation, which corresponds to the widely used standard Born-Oppenheimer approximation \cite{Born:1927}.
	
	While corrections to the ordinary static potential up to order $(1 /m_Q)^2$ have been studied using first principles lattice gauge theory (see e.g.\ Refs.\ \cite{Bali:1997am,Koma:2006si,Koma:2006fw,Eichberg:2023trq}), the next-to-leading order (i.e.\ order $(1 /m_Q)^1$) corrections to hybrid static potentials have not yet been computed using lattice techniques.
	Unlike ordinary quarkonium, where spin effects are suppressed by $(1 /m_Q)^2$, spin-dependent contributions to heavy hybrid mesons such as spin-dependent potentials and hybrid-quarkonium mixing potentials appear already at order $(1 /m_Q)^1$ \cite{Oncala:2017hop,Soto:2017one,Brambilla:2018pyn,Brambilla:2019jfi}.
	Thus, these contributions are even more important for precise predictions of heavy hybrid meson spectra.

	Spin effects in heavy hybrid mesons and hybrid-quarkonium mixing at order $(1 /m_Q)^1$ were investigated some time ago with lattice Non-Relativistic QCD (NRQCD) \cite{Burch:2001nk,Burch:2003zf} and more recently full lattice QCD studies of charmonium hybrids \cite{HadronSpectrum:2012gic,Cheung:2016bym} and bottomonium hybrids \cite{Gayer:2024akw} have been carried out.
	For small quark-antiquark separations, spin-dependent corrections to hybrid potentials have been explored in weakly-coupled potential NRQCD (pNRQCD) up to order $(1 /m_Q)^2$ in Refs.\ \cite{Brambilla:2018pyn,Brambilla:2019jfi}.
	Each of these corrections can be written as a sum of a $r$-dependent part, which was calculated perturbatively, and a non-perturbative part, which was expressed in a multipole expansion in terms of  $r$-independent gluonic correlators.
	For large separations, the form of the potential corrections can be described with QCD effective string theory \cite{Luscher:2002qv,Luscher:2004ib}.
	This was used in Refs.~\cite{Oncala:2017hop,Soto:2023lbh} to parametrize hybrid potentials at order $(1 /m_Q)^1$ employing \mbox{pNRQCD} for small separations and QCD effective string theory for large separations.
	Investigations of spin splitting and mixing effects in heavy hybrid meson spectra have been carried out in Refs.\ \cite{Oncala:2017hop,Brambilla:2018pyn,Brambilla:2019jfi,Soto:2023lbh}, but each of these works involves unknown parameters, which can e.g.\ be computed using lattice gauge theory.
	
 	Recently, expressions for hybrid spin-dependent and hybrid-quarkonium mixing potentials at order $(1 /m_Q)^1$ have been derived in the BOEFT framework without assumptions or restrictions on the quark-antiquark separation \cite{Oncala:2017hop,Soto:2020xpm}.
	The potentials are given by integral expressions including generalized Wilson loops with a chromomagnetic field insertion along one of the temporal lines and suitable hybrid creation operators replacing the spatial lines.
	Through straightforward analytic integration, the potentials can be related to matrix elements that are suited for evaluation with lattice gauge theory.
	The results presented in this work are the first lattice gauge theory results for the hybrid spin-dependent and hybrid-quarkonium mixing potentials at order $(1 /m_Q)^1$.
	Even though limited to a single lattice spacing $a = 0.060 \, \text{fm}$, they provide important insights into the $r$-dependence of these potentials. 
	Similar future computations at different lattice spacings and a subsequent continuum extrapolation would lead to a fully rigorous first principles prediction of $(1 /m_Q)^1$ hybrid spin-dependent and hybrid-quarkonium mixing potentials, which could be used for refined BOEFT predictions of heavy hybrid meson spectra.

	This paper is structured in the following way.
	Section~\ref{sec:BOEFTpotentials} summarizes the BOEFT as developed in Refs.\ \cite{Berwein:2015vca,Oncala:2017hop,Brambilla:2017uyf,Brambilla:2018pyn,Brambilla:2019jfi,Soto:2020xpm}.
	After specializing the equations to heavy hybrid mesons with gluon spin quantum numbers $\kappa^{PC} = 1^{+-}$, it becomes obvious that there are four relevant $(1/m_Q)^1$ potentials, $V^{sa}_{11}(r)$, $V^{sb}_{10}(r)$, $V^\text{mix}_{\Sigma_u^-}(r)$ and $V^\text{mix}_{\Pi_u}(r)$.
	In Section~\ref{sec:latticesetup} we relate these potentials to matrix elements in SU(3) gauge theory.
	We also discuss our lattice setup and techniques, including details of the generated gauge link ensemble, suitable lattice creation operators and the discretization of the chromomagnetic field operator, correlation functions for the extraction of the matrix elements and the application of gradient flow.
	In Section~\ref{sec:results} we present our numerical results, in particular results for the four potentials $V^{sa}_{11}(r)$, $V^{sb}_{10}(r)$, $V^\text{mix}_{\Sigma_u^-}(r)$ and $V^\text{mix}_{\Pi_u}(r)$ at lattice spacing $a = 0.060 \, \text{fm}$.
	We conclude in Section~\ref{sec:conclusion}.


\newpage
	
	\section{Born-Oppenheimer Effective Field Theory for heavy hybrid mesons with gluon spin $\kappa^{PC} = 1^{+-}$}\label{sec:BOEFTpotentials}
	In this section we specialize the BOEFT for heavy hybrid mesons and ordinary quarkonium, which was developed in Refs.\ \cite{Berwein:2015vca,Oncala:2017hop,Brambilla:2017uyf,Brambilla:2018pyn,Brambilla:2019jfi,Soto:2020xpm}, to hybrid mesons with gluon spin quantum numbers $\kappa^{PC} = 1^{+-}$.
	The corresponding Lagrangian can be organized in powers of the inverse heavy quark mass $1/m_Q$. It contains a priori unknown potentials, which are, however, suited for computations with lattice gauge theory.
	
	At leading order (i.e.\ order $(1/m_Q)^0$) the relevant potentials are the hybrid static potentials with quantum numbers $\Lambda_\eta^\epsilon = \Sigma_u^- , \Pi_u$ (for details see Section~\ref{sec:LO}).
	They were recently recomputed with lattice gauge theory with unprecedented precision and resolution \cite{Schlosser:2021wnr} (for earlier computations see Refs.\ \cite{
		Griffiths:1983ah,
		Campbell:1984fe,
		Campbell:1987nv,
		Perantonis1989StaticPF,
		Michael:1990az,
		Perantonis:1990dy,
		Juge:1997nc,
		Peardon:1997jr,
		Juge:1997ir,
		Morningstar:1998xh,
		Michael:1998tr,
		Michael:1999ge,
		Juge:1999ie,
		Juge:1999aw,
		Bali:2000vr,
		Morningstar:2001nu,
		Juge:2002br,
		Michael:2003ai,
		Michael:2003xg,
		Bali:2003jq,
		Juge:2003ge,
		Wolf:2014tta,
		Bicudo:2018jbb,
		Capitani:2018rox}).
	
	At next-to-leading order (i.e.\ order $(1/m_Q)^1$) we consider spin-dependent interactions, which are responsible for the hyperfine splitting in spectra of heavy hybrid mesons.
	There are two corresponding potentials, $V^{sa}_{11}(r)$ and $V^{sb}_{10}(r)$ ($r$ denotes the heavy quark-antiquark separation)~\cite{Soto:2020xpm,Soto:2023lbh}.
	Neither of them has yet been computed from first principles.
	We also consider mixing of heavy hybrid mesons with ordinary quarkonium at order $(1/m_Q)^1$. In addition to the well-known ordinary static potential with quantum numbers $\Lambda_\eta^\epsilon = \Sigma_g^+$, which is needed for the quarkonium Lagrangian at order $(1/m_Q)^0$, there are two unknown mixing potentials, $V^\text{mix}_{\Sigma_u^-}(r)$ and $V^\text{mix}_{\Pi_u}(r)$~\cite{Oncala:2017hop}.
	
	As already stated in the introduction, it is the main goal of this work to carry out the first lattice gauge theory computation of these four $(1/m_Q)^1$ potentials, $V^{sa}_{11}(r)$, $V^{sb}_{10}(r)$, $V^\text{mix}_{\Sigma_u^-}(r)$ and $V^\text{mix}_{\Pi_u}(r)$. 
	The brief discussion and summary of the BOEFT in this section is intended to motivate our lattice computation discussed in Section~\ref{sec:latticesetup} and Section~\ref{sec:results}.
	Moreover, it should provide readers, who are unfamiliar with the hybrid-quarkonium BOEFT, some background, where these potentials are needed and why they are of significant interest.
	For further details on the BOEFT we recommend Refs.\ \cite{Berwein:2015vca,Oncala:2017hop,Brambilla:2017uyf,Brambilla:2018pyn,Brambilla:2019jfi,Soto:2020xpm,Berwein:2024ztx}.
	
	
		\subsection{\label{sec:spin_potentials}Lagrangian for heavy hybrid mesons}
		The BOEFT Lagrangian for heavy hybrid mesons with gluon quantum numbers $\kappa^{PC} = 1^{+-}$ including corrections due to the finite mass of the heavy quarks, but without mixing with ordinary quarkonium, is given by
		\begin{align}\label{eq:L_hybrids} 
			\mathcal{L} = \Psi^{\dagger}_{1^{+-}} \Big(i\partial_t - h_{1^{+-}}(\mathbf{r})\Big) \Psi_{1^{+-}}
		\end{align}
		with
		\begin{align}
			h_{1^{+-}}(\mathbf{r}) = -\frac{\Delta_\mathbf{r}}{m_Q} + V^{(0)}_{1^{+-}}(\mathbf{r}) + \frac{1}{m_Q}  V^{(1)}_{1^{+-}}(\mathbf{r},\textbf{p}) + \mathcal{O}((1/m_Q)^2)
		\end{align}
		(see Refs.\ \cite{Soto:2020xpm,Soto:2023lbh}).
		The field $\Psi_{1^{+-}} \equiv \Psi^{nA}_{1^{+-}}$ has twelve components
		representing the three possible orientations for gluon spin $\kappa = 1$ (labeled by the index $n$) and the four possible spin orientations of the heavy quark-antiquark pair (labeled by the index $A$).
		Correspondingly, the potentials $V^{(0)}_{1^{+-}}(\mathbf{r}) \equiv V^{(0) \, n' A';n A}_{1^{+-}}(\mathbf{r})$ and $V^{(1)}_{1^{+-}}(\mathbf{r},\textbf{p}) \equiv V^{(1) \, n' A';n A}_{1^{+-}}(\mathbf{r},\textbf{p})$ are $12 \times 12$ matrices.
		
		
			\subsubsection{Order $(1 /m_Q)^0$}\label{sec:LO}
			The leading order potential $V^{(0)}_{1^{+-}}(\mathbf{r})$ can be expressed in terms of hybrid static potentials $V_{\Lambda_{\eta}^{\epsilon}}(r)$,
			\begin{equation}\label{eq:staticpotential_decomposition} 
				V^{(0) \, n' A';n A}_{1^{+-}}(\mathbf{r}) = \delta^{A' A} \sum_{\Lambda_\eta^\epsilon = \Sigma_u^- , \Pi_u} V_{\Lambda_\eta^\epsilon}(r) \mathcal{P}^{n' n}_{1 \Lambda} .
			\end{equation}
			Hybrid static potentials are labeled by quantum numbers $\Lambda_\eta^\epsilon$, which classify representations of the dihedral group $D_{\infty h}$:
			\begin{itemize}
				\item $\Lambda = \Sigma (=0), \Pi (=1), \Delta (=2), \ldots$ denotes the absolute value of the total angular momentum with respect to the quark-antiquark separation axis, i.e.\ $\Lambda$ is a non-negative integer (for the computations discussed in Section~\ref{sec:latticesetup} and Section~\ref{sec:results} we choose the $z$-axis as separation axis).
				
				\item $\eta = g (=+), u (=-)$ describes the even or odd behavior with respect to the combined parity and charge conjugation transformation $\mathcal{P} \circ \mathcal{C}$.
				
				\item $\epsilon = +,-$ is the eigenvalue of the reflection along an axis perpendicular to the quark-antiquark separation axis (for the computations discussed in Section~\ref{sec:latticesetup} and Section~\ref{sec:results} we choose the $x$-axis).
				For $\Lambda \ge 1$ hybrid potentials are degenerate with respect to $\epsilon$ and, thus, $\epsilon$ is typically omitted.
				For example $V_{\Pi_u}(r) \equiv V_{\Pi_u^+}(r) = V_{\Pi_u^-}(r)$, which was used to simplify Eqs.\ (\ref{eq:staticpotential_decomposition}) to (\ref{eq:staticpotential_decomposition1_simplified}).
				Note, however, that	states with quantum numbers $\Pi_u^+$ and $\Pi_u^-$ are orthogonal and need to be differentiated, when computing matrix elements.
			\end{itemize}
			The $3 \times 3$ matrices $\mathcal{P}_{1 \Lambda} \equiv \mathcal{P}^{n' n}_{1 \Lambda}$ relate the three gluon spin components of the field $\Psi_{1^{+-}}$ to the hybrid static potentials $V_{\Sigma_u^-}(r)$ and $V_{\Pi_u}(r)$.
			In this paper we work with fields $\Psi^{nA}_{1^{+-}}$, where the gluon spin index $n = x,y,z$ labels eigenstates of the three Cartesian $x$-, $y$- and $z$-components of the gluon spin-$1$ operator.
			Then
			\begin{equation}\label{eq:projectors} 
				\mathcal{P}_{1 \Sigma} = \mathcal{P}_{1 0} = \mathbf{e}_r \otimes \mathbf{e}_r \quad , \quad \mathcal{P}_{1 \Pi} = \mathcal{P}_{1 1} = 1 - \mathbf{e}_r \otimes \mathbf{e}_r
			\end{equation}
			or, equivalently,
			\begin{equation}
				V^{(0) n' A';n A}_{1^{+-}}(\mathbf{r}) = \delta^{A' A} \Big(V_{\Sigma_u^-}(r) \Big(\mathbf{e}_r \otimes \mathbf{e}_r\Big)^{n' n} + V_{\Pi_u}(r) \Big(1 - \mathbf{e}_r \otimes \mathbf{e}_r\Big)^{n' n}\Big) . \label{eq:staticpotential_decomposition1_simplified}
			\end{equation}
			Note that at order $(1 /m_Q)^0$ the spin of the heavy quarks is irrelevant, as indicated by $\delta^{A' A}$.
			
		
			\subsubsection{Order $(1 /m_Q)^1$}\label{sec:NLO}
			At order $(1 /m_Q)^1$ there are both spin-dependent and spin-independent contributions to the potential $V^{(1)}_{1^{+-}}(\mathbf{r},\textbf{p})$,
			\begin{equation}
				V^{(1)}_{1^{+-}}(\mathbf{r},\textbf{p}) = V^{(1),\text{SD}}_{1^{+-}}(\mathbf{r}) + V^{(1),\text{SI}}_{1^{+-}}(\mathbf{r},\textbf{p}) .
			\end{equation}
			We do not consider the spin-independent part $V^{(1),\text{SI}}_{1^{+-}}(\mathbf{r},\textbf{p})$ in this work, i.e.\ from now on
			\begin{equation}
				V^{(1)}_{1^{+-}}(\mathbf{r},\textbf{p}) = V^{(1),\text{SD}}_{1^{+-}}(\mathbf{r}) .
			\end{equation}
			The reason, why we neglect $V^{(1),\text{SI}}_{1^{+-}}(\mathbf{r},\textbf{p})$, is that matching equations are mostly not available (only an angular momentum-dependent potential $V^l_{\Lambda \Lambda^\prime}(r)$ has been derived; see Ref.\ \cite{Soto:2020xpm} for details).
			Note that BOEFT computations of heavy hybrid meson masses with spin-independent terms neglected, as sketched in Section~\ref{sec:SE}, are still of interest, because those terms do not affect the predicted spin splittings.
			
			The spin-dependent potential at order $(1/m_Q)^1$ is given in Refs.\ \cite{Brambilla:2018pyn,Brambilla:2019jfi} and expressed in a different notation in Refs.\ \cite{Soto:2020xpm,Soto:2023lbh} for arbitrary gluon spin quantum numbers $\kappa^{PC}$.
			In this paper we work with the latter notation (see Eq.\ (4) in Ref.\ \cite{Soto:2020xpm} and Eq.\ (1) in Ref.\ \cite{Soto:2023lbh}). For $\kappa^{PC} = 1^{+-}$ and with all indices explicitly written this potential is
			\begin{eqnarray}
				\nonumber & & \hspace{-0.7cm} V^{(1),\text{SD} \, n' A';n A}_{1^{+-}}(\mathbf{r}) = \mathcal{P}^{n'j}_{1 1} V^{sa}_{11}(r) \Big((\bm{S}^p_{Q \bar{Q}})^{A' A} \mathcal{P}^{pq}_{1 0} (\bm{S}^q_1)^{j k}\Big) \mathcal{P}^{kn}_{1 1} \\
				\label{EQN_V1} & & \hspace{0.7cm} + 
				\mathcal{P}^{n'j}_{1 1} V^{sb}_{10}(r) \Big((\bm{S}^p_{Q \bar{Q}})^{A' A} \mathcal{P}^{pq}_{1 1} (\bm{S}^q_1)^{j k}\Big) \mathcal{P}^{kn}_{1 0} + 
				\mathcal{P}^{n'j}_{1 0} V^{sb}_{01}(r) \Big((\bm{S}^p_{Q \bar{Q}})^{A' A} \mathcal{P}^{pq}_{1 1} (\bm{S}^q_1)^{j k}\Big) \mathcal{P}^{kn}_{1 1} ,
			\end{eqnarray}
			where $(\bm{S}^q_1)^{j k} = - i \epsilon^{q j k}$, $q = x,y,z$ denote the three Cartesian $x$-, $y$- and $z$-components of the gluon spin-$1$ operator. 
			$(\bm{S}^p_{Q \bar{Q}})^{A' A}$, $p = x,y,z$ are the three Cartesian $x$-, $y$- and $z$-components of the heavy quark spin operator
			\footnote{
				For Eq.\ (\ref{EQN_V1}) it is not necessary to fix, which heavy spin combinations correspond to the indices $A, A' = 0,1,2,3$. In Section~\ref{sec:mixing} this is different:
				$A, A' = 0$ will represent the $S_{Q \bar{Q}} = 0$ singlet and $A, A' = 1,2,3$ will represent the $S_{Q \bar{Q}} = 1$ triplet.}.
			It can be shown  as a result of time-reversal symmetry and hermiticity that $V^{sb}_{10}(r) = V^{sb}_{01}(r)$, i.e.\ there are only two independent potentials on the right hand side of Eq.\ (\ref{EQN_V1}), $V^{sa}_{11}(r)$ and $V^{sb}_{10}(r)$.

			Notice that $V^{(1),\text{SD}}_{1^{+-}}(\mathbf{r})$ is proportional to $(\bm{S}^p_{Q \bar{Q}})$.
			Thus, for heavy quark spin $S_{Q \bar{Q}} = 0$ the spin-dependent potential vanishes, i.e.\ $V^{(1),\text{SD}}_{1^{+-}}(\mathbf{r}) = 0$.
			We also note that $V^{sa}_{11}(r)$ is multiplied to the vector-times-projector-times-vector term $\bm{S}_{Q \bar{Q}} \cdot \mathcal{P}_{1 0} \cdot \bm{S}_1 = \bm{S}_{Q \bar{Q}} \cdot (\mathbf{e}_r \otimes \mathbf{e}_r) \cdot \bm{S}_1$, which implies that $V^{sa}_{11}(r)$ is most relevant, when both the gluon spin and the heavy quark spin are aligned along the quark-antiquark separation axis.
			Similarly, $V^{sb}_{10}(r)$ is multiplied to the term $\bm{S}_{Q \bar{Q}} \cdot \mathcal{P}_{1 1} \cdot \bm{S}_1 = \bm{S}_{Q \bar{Q}} \cdot (1 - \mathbf{e}_r \otimes \mathbf{e}_r) \cdot \bm{S}_1$, i.e.\ it is most important, when both the gluon spin and the heavy quark spin are orthogonal to the quark-antiquark separation axis.
	
		
		\subsection{\label{sec:mixing}Lagrangian for hybrid-quarkonium mixing}
		The BOEFT Lagrangian for both ordinary quarkonium and heavy hybrid mesons with gluon spin quantum numbers $\kappa^{PC} = 1^{+-}$ including hybrid-quarkonium mixing at order $(1/m_Q)^1$ is given in Ref.\ \cite{Oncala:2017hop},
		\begin{eqnarray}
			\nonumber & & \hspace{-0.7cm} \mathcal{L} = \underbrace{\tr\Big(S^\dagger \Big(i \partial_t - h_S(\mathbf{r})\Big) S\Big)}_{= \mathcal{L}_\text{quarkonium}} + \underbrace{\tr\Big(H^{n' \dagger} \Big(i \delta^{n' n} \partial_t - h_H^{n' n}(\mathbf{r})\Big) H^n\Big)}_{= \mathcal{L}_\text{hybrids}} \\
			\label{EQN_L_mixing} & & \hspace{0.7cm} + \underbrace{\tr\Big(S^\dagger V^{n' n}_\text{mix}(\mathbf{r}) \{\sigma^{n'} , H^n\} + \text{H.c.}\Big)}_{= \mathcal{L}_\text{mixing}} ,
		\end{eqnarray}
		where
		\begin{eqnarray}
			 & & \hspace{-0.7cm} h_S(\mathbf{r}) = -\frac{\Delta_\mathbf{r}}{m_Q} + V_{\Sigma_g^+}(r) \\
			\nonumber & & \hspace{-0.7cm} h_{H}^{n' n}(\mathbf{r}) = -\frac{\Delta_\mathbf{r}}{m_Q} \delta^{n' n} + \sum_{\Lambda_\eta^\epsilon = \Sigma_u^- , \Pi_u} V_{\Lambda_\eta^\epsilon}(r) \mathcal{P}^{n' n}_{1 \Lambda} = \\
			 & & = -\frac{\Delta_\mathbf{r}}{m_Q} \delta^{n' n} + V_{\Sigma_u^-}(r) \Big(\mathbf{e}_r \otimes \mathbf{e}_r\Big)^{n' n} + V_{\Pi_u}(r) \Big(1 - \mathbf{e}_r \otimes \mathbf{e}_r\Big)^{n' n} \\
			\label{eq:V_mix} & & \hspace{-0.7cm} V^{n' n}_\text{mix}(\mathbf{r}) = \sum_{\Lambda_\eta^\epsilon = \Sigma_u^- , \Pi_u} V^{\text{mix}}_{\Lambda_\eta^\epsilon}(r) \mathcal{P}^{n' n}_{1 \Lambda} = V^{\text{mix}}_{\Sigma_u^-}(r) \Big(\mathbf{e}_r \otimes \mathbf{e}_r\Big)^{n' n} + V^{\text{mix}}_{\Pi_u}(r) \Big(1 - \mathbf{e}_r \otimes \mathbf{e}_r\Big)^{n' n}
		\end{eqnarray}
		and $\text{H.c.}$ denotes the hermitian conjugate. $S = S_0 + \sigma^j S_1^j$ is the ordinary quarkonium field and $H^n = H^n_0 + \sigma^j H_1^{n j}$ the hybrid quarkonium field.
		The lower indices $0$ and $1$ denote heavy quark spin $S_{Q \bar{Q}} = 0$ and $S_{Q \bar{Q}} = 1$, respectively.
		The upper index $n$ is the gluon spin index, which labels eigenstates of the three Cartesian $x$-, $y$- and $z$-components of the gluon spin-$1$ operator.
		Each term inside of $\tr(\ldots)$ in Eq.\ (\ref{EQN_L_mixing}) is a $2 \times 2$ matrix and $\tr(\ldots)$ denotes the trace with respect to these $2\times 2$ matrices.
		$h_S$ and $h_H$ are the order $(1/m_Q)^0$ Hamiltonians for ordinary quarkonium and hybrid quarkonium, respectively.
		The term $\mathcal{L}_\text{mixing}$ in the Lagrangian (\ref{EQN_L_mixing}) describes hybrid-quarkonium mixing, which is related to the heavy quark spin.
		It is suppressed and proportional to $1/m_Q$ (see Eqs.\ (\ref{eq:VmixPi}) and (\ref{eq:VmixSigma}) and Ref.\ \cite{Oncala:2017hop} for details).
					
		The notation for the fields from Ref.\ \cite{Oncala:2017hop} (used in Eqs.\ (\ref{EQN_L_mixing}) to (\ref{eq:V_mix})) is somewhat different from the notation from Refs.\ \cite{Soto:2020xpm,Soto:2023lbh} (used in Section~\ref{sec:spin_potentials}).
		It can, however, easily be translated:
		\begin{itemize}
			\item $\Psi^{n A}_{1^{+-}} = H_0^n$, if $A = 0$,
			\item $\Psi^{n A}_{1^{+-}} = H_1^{n j}$, if $A = j$.
		\end{itemize}
		The physical interpretation of the four heavy quark spin components, labeled by $A = 0,1,2,3$ in $\Psi^{n A}_{1^{+-}}$, is then, however, fixed: 
		$A = 0$ refers to heavy quark-antiquark spin $S_{Q \bar Q} = 0$ and $A = j$ with $j = 1,2,3 \equiv x,y,z$ to the eigenstates of the three Cartesian $x$-, $y$- and $z$-components of the heavy quark spin-1 operator, i.e.\ to $S_{Q \bar Q} = 1$.
		The term $\mathcal{L}_\text{hybrids}$ in Eq.\ (\ref{EQN_L_mixing}) is then identical to the Lagrangian (\ref{eq:L_hybrids}) restricted to order $(1/m_Q)^0$, i.e.\ with the potential (\ref{eq:staticpotential_decomposition1_simplified}), but without spin-dependent potentials.
		The Lagrangian for ordinary quarkonium can also be written in the style of Section~\ref{sec:spin_potentials},
		\begin{equation}
			\mathcal{L}_\text{quarkonium} = \Psi^{\dagger}_{0^{++}} \Big(i\partial_t - h_{0^{++}}(\mathbf{r})\Big) \Psi_{0^{++}}
		\end{equation}
		with $h_{0^{++}}(\mathbf{r}) = h_S(\mathbf{r})$ and a four-component field $\Psi_{0^{++}} \equiv \Psi^A_{0^{++}}$ related to $S$ via
		\begin{itemize}
			\item $\Psi^A_{0^{++}} = S_0$, if $A = 0$,
			\item $\Psi^A_{0^{++}} = S_1^j$, if $A = j$.
		\end{itemize}
		Finally, the mixing term in the Lagrangian (\ref{EQN_L_mixing}) can be rewritten and expressed in terms of the $\Psi$ fields,
		\begin{equation}
			\mathcal{L}_\text{mixing} = 2 V^{n' n}_\text{mix} \Big(S_1^{n' \dagger} H_0^n + S_0^\dagger H_1^{n n'} + \text{H.c.}\Big) = 2 V^{n' n}_\text{mix} \Big(
			\Psi^{n' \dagger}_{0^{++}} \Psi^{n 0}_{1^{+-}} +
			\Psi^{0 \dagger}_{0^{++}} \Psi^{n n'}_{1^{+-}} + \text{H.c.}\Big) .
		\end{equation}
		
		
		\subsection{\label{sec:SE}Coupled channel Schr\"odinger equations}
		From the Lagrangians discussed in Section~\ref{sec:spin_potentials} and Section~\ref{sec:mixing} one can derive coupled channel Schr\"o\-dinger equations for the radial coordinate of the quark-antiquark separation.
		Solving these equations leads to predictions of spectra and properties of heavy hybrid mesons.
		These Schr\"odinger equations include the potentials $V^{sa}_{11}(r)$, $V^{sb}_{10}(r)$, $V^\text{mix}_{\Sigma_u^-}(r)$ and $V^\text{mix}_{\Pi_u}(r)$, which we compute with lattice gauge theory in Section~\ref{sec:latticesetup} and Section~\ref{sec:results}.
		Hence, they are an important ingredient for BOEFT predictions of heavy hybrid meson masses. 
		
		At order $(1/m_Q)^0$, Schr\"odinger equations for heavy hybrid mesons with gluon spin quantum numbers $\kappa^{PC} = 1^{+-}$ are either single channel equations or $2 \times 2$ coupled channel equations \cite{Berwein:2015vca,Oncala:2017hop}.
		They contain the hybrid static potentials $V_{\Sigma_u^-}(r)$ and $V_{\Pi_u}(r)$ as evident from Eq.\ (\ref{eq:staticpotential_decomposition1_simplified}).
		These leading order equations do not contain the spin of the heavy quark-antiquark pair and, consequently, resulting energy levels are degenerate with respect to $S_{Q\bar Q}$.
		
		The spin-dependent potential $V_{1^{+-}}^{(1),SD}(\mathbf{r})$ (see Eq.\ (\ref{EQN_V1})) appearing at order $(1/m_Q)^1$ is responsible for the hyperfine splitting in heavy hybrid meson spectra.
		It leads to coupled channel Schr\"odinger equations, which have a $3 \times 3$, $7 \times 7$ and $9 \times 9$ matrix structure for total angular momentum of the heavy hybrid meson $\mathcal{J} = 0$, $\mathcal{J} = 1$ and $\mathcal{J} > 1$, respectively \cite{Soto:2023lbh}.
		$V_{1^{+-}}^{(1),SD}(\mathbf{r})$ can be expressed in terms of two radially symmetric potentials $V^{sa}_{11}(r)$ and $V^{sb}_{10}(r)$, which have not yet been computed from first principles with lattice gauge theory.
		There are, however, predictions from pNRQCD and QCD effective string theory for the small and large $r$ behavior of $V^{sa}_{11}(r)$ and $V^{sb}_{10}(r)$ \cite{Brambilla:2018pyn,Brambilla:2019jfi,Soto:2023lbh}.
		They were used in these references to determine the unknown potentials and to study the hyperfine splitting by solving the previously mentioned coupled Schrödinger equations.
		
		The potential $V_\text{mix}(\mathbf{r})$ (see Eq.\ (\ref{eq:V_mix})), also appearing at order $(1/m_Q)^1$, causes a mixing of ordinary quarkonium and heavy hybrid mesons.
		When neglecting the spin-dependent potential $V_{1^{+-}}^{(1),SD}(\mathbf{r})$, it leads to $2 \times 2$, $4 \times 4$ and $6 \times 6$ coupled channel Schr\"odinger equations\cite{Oncala:2017hop}.
		The mixing potential can be expressed in terms of two radially symmetric potentials, $V^\text{mix}_{\Sigma_u^-}(r)$ and $V^\text{mix}_{\Pi_u}(r)$, which have as well not yet been computed from first principles with lattice gauge theory.
		Again, there are predictions based on pNRQCD and QCD effective string theory for the small and large $r$ behavior of $V^\text{mix}_{\Sigma_u^-}(r)$ and $V^\text{mix}_{\Pi_u}(r)$, which were used for modeling these unknown potentials to study hybrid-quarkonium mixing in Ref.\ \cite{Oncala:2017hop}.
		
		An illustrative and representative example for all the coupled channel Schr\"odinger equations previously mentioned in this subsection is the $2 \times 2$ hybrid-quarkonium mixing equation
		\begin{equation}
			\bigg(-\frac{1}{m_Q} \frac{d^2}{d r^2} + \frac{\j (\j + 1)}{m_Q r^2} + \bigg(\begin{array}{cc}
			  V_{\Sigma_g^+}(r) & 2 V^\text{mix}_{\Pi_u}(r) \\
			  2 V^\text{mix}_{\Pi_u}(r) & V_{\Pi_u}(r)
			\end{array}\bigg)\bigg)
			\begin{pmatrix}
			  S_{1 \mathcal{J} \mathcal{M}}^0(r) \\
			  P_{0 \mathcal{J} \mathcal{M}}^0(r)
			\end{pmatrix}
			= E
			\begin{pmatrix} 
			  S_{1 \mathcal{J} \mathcal{M}}^0(r) \\
			  P_{0 \mathcal{J} \mathcal{M}}^0(r)
			\end{pmatrix}
		\end{equation}
		valid for total angular momentum $\mathcal{J} \neq 0$ of both quarkonium and the heavy hybrid meson.
		In this particular case it is rather obvious that the mixing potential $V^\text{mix}_{\Pi_u}(r)$, appearing in the off-diagonal elements of the potential matrix, couples the radial wave functions $S_{1\mathcal{J}\mathcal{M}}^0$ and $P_{0\mathcal{J}\mathcal{M}}^0$.
		$S_{1\mathcal{J}\mathcal{M}}^0$ represents quarkonium with $S_{Q \bar Q} = 1$ (the upper index $0$ indicates that $L=J$, where $J$ is the quark orbital angular momentum plus the gluon total angular momentum), while $P_{0\mathcal{J}\mathcal{M}}^0$ represents a heavy hybrid meson with $S_{Q \bar Q} = 0$.
		

	\newpage
		
		
	\section{Lattice setup and techniques}\label{sec:latticesetup}
	
		\subsection{\label{SEC_ensemble}Lattice gauge link ensemble}
		To complement our recent precision computations of the hybrid static potentials $V_{\Sigma_u^-}(r)$ and $V_{\Pi_u}(r)$ discussed in detail in Ref.\ \cite{Schlosser:2021wnr}, we use for this exploratory study one of the four SU(3) lattice gauge link ensembles from our previous work \cite{Schlosser:2021wnr,Herr:2023xwg}, ensemble $B$ with lattice spacing $a = 0.060 \, \text{fm}$.
		This ensemble was generated with the standard Wilson plaquette action for SU(3) gauge theory without dynamical quarks using the CL2QCD software package\cite{Philipsen:2014mra}.
		
		The gauge coupling is $\beta=6.284$.
		We related the corresponding lattice spacing $a$ to the Sommer scale $r_0$ via a parametrization of $\ln(a/r_0)$ provided in Ref.\ \cite{Necco:2001xg}.
		Then we introduced physical units by setting $r_0 = 0.5 \, \text{fm}$, which is a simple and common choice in pure gauge theory.
		A computation at the corresponding intermediate lattice spacing $a = 0.060 \, \text{fm}$ \footnote{The ensembles used in Ref.\ \cite{Schlosser:2021wnr} cover lattice spacings in the range $0.040 \, \text{fm} \leq a \leq 0.093 \, \text{fm}$.} provides potentials $V^{sa}_{11}(r)$, $V^{sb}_{10}(r)$, $V^{\rm mix}_{\Sigma_u^-}(r)$ and $V^\text{mix}_{\Pi_u}(r)$ also at intermediate quark-antiquark separations.
		Thus, it seems to be a good starting point for future computations of these $(1 / m_Q)^1$ potentials with several larger and smaller lattice spacings.
		The lattice volume is $(20a)^3 \times 40a = (1.2 \, \text{fm})^3 \times 2.4 \,\text{fm}$.
		This is sufficiently large to neglect finite volume corrections (see Ref.\ \cite{Schlosser:2021wnr} for a detailed investigation and discussion).
		
		The ensemble was generated by $N_{\text{sim}}=2$ independent Monte Carlo simulations, where each simulation comprises $N_{\text{total}}=85000$ updates.
		An update is composed of a heatbath sweep and $N_{\text{or}}=12$ overrelaxation sweeps.
		The first $N_{\text{therm}}=20000$ thermalization updates were discarded.
		Then gauge link configurations separated by $N_{\text{sep}}=100$ updates were used to evaluate correlation functions.
		Statistical errors were determined using the jackknife method, where we combined the $N_{\text{meas}}=1300$ gauge link configurations used for measurements to $260$ reduced jackknife bins.
		Further details concerning our data analysis are described in Appendix B of Ref.\ \cite{Schlosser:2021wnr}.
			
		
		\subsection{\label{SEC_3_2}Relating the $(1 / m_Q)^1$ potentials to matrix elements in SU(3) gauge theory}
		By matching the BOEFT to NRQCD up to order $(1/m_Q)^1$, the potentials $V^{sa}_{11}(r)$, $V^{sb}_{10}(r)$, $V^\text{mix}_{\Sigma_u^-}(r)$ and $V^\text{mix}_{\Pi_u}(r)$ can be expressed in terms of generalized Wilson loops
		\footnote{
			With ``generalized Wilson loops'' we refer to Wilson loops with chromomagnetic field insertions $B_j$, $j=x,y,z$ on one of the temporal lines, and where the spatial parallel transporters are not just straight lines, but more complicated structures exciting gluons with definite $\Lambda_\eta^\epsilon$ quantum numbers.}.
		This was done in Ref.\ \cite{Soto:2020xpm} for the spin-dependent potentials $V^{sa}_{11}(r)$ and $V^{sb}_{10}(r)$ and in Ref.\ \cite{Oncala:2017hop} for the mixing potentials $V^\text{mix}_{\Sigma_u^-}(r)$ and $V^\text{mix}_{\Pi_u}(r)$.
		
		The resulting expressions (Eqs.\ (33) and (34) in Ref.\ \cite{Soto:2020xpm} and Eqs.\ (28) and (29) in Ref.\ \cite{Oncala:2017hop}) are integrals over the temporal position of a chromomagnetic field insertion in a generalized Wilson loop. These expressions are, however, not practical for lattice computations.
		We have, thus, solved the integrals analytically, which is straightforward after inserting sums over complete sets of energy eigenstates and taking the limit of infinite temporal separation.
		This led to rather simple relations of the $(1/m_Q)^1$ potentials and matrix elements, where the bra's and the ket's are ground states that either correspond to the ordinary static potential with quantum numbers $\Sigma_g^+$ or to the hybrid static potentials with quantum numbers $\Sigma_u^-$ and $\Pi_u$,
		\begin{eqnarray}
			\label{eq:Vsa} & & \hspace{-0.7cm} V^{sa}_{11}(r) = i gc_F \bra{0 , \Pi_u^-} B_z(-r/2) \ket{0 , \Pi_u^+} (r) \\
			\label{eq:Vsb} & & \hspace{-0.7cm} V^{sb}_{10}(r) = i gc_F \bra{0 , \Sigma_u^-} B_y(-r/2) \ket{0 , \Pi_u^+}(r) = i gc_F \bra{0 , \Sigma_u^-} B_x(-r/2) \ket{0 , \Pi_u^-} (r) \\
			\label{eq:VmixPi} & & \hspace{-0.7cm} V^\text{mix}_{\Pi_u}(r) = \frac{i gc_F}{2 m_Q} \bra{0 , \Sigma_g^+} B_x(-r/2) \ket{0 , \Pi_u^+}(r) = -\frac{i gc_F}{2 m_Q} \bra{0 , \Sigma_g^+} B_y(-r/2) \ket{0 , \Pi_u^-} (r) \\
			\label{eq:VmixSigma} & & \hspace{-0.7cm} V^\text{mix}_{\Sigma_u^-}(r) = \frac{i gc_F}{2 m_Q} \bra{0 , \Sigma_g^+} B_z(-r/2) \ket{0,{\Sigma_u^-}} (r) .
		\end{eqnarray}
		In detail, $| 0,\Lambda_\eta^\epsilon \rangle$ denotes the ground state of a static quark at position $(0,0,+r/2)$ and a static antiquark at position $(0,0,-r/2)$, where the connecting flux tube has quantum numbers $\Lambda_\eta^\epsilon$.
		Since for all matrix elements appearing in Eqs.\ (\ref{eq:Vsa}) to (\ref{eq:VmixSigma}) the ground states represented by the bra and the ket are from different $\Lambda_\eta^\epsilon$ sectors, their relative phases are not arbitrary.
		The ground states are, thus, defined as
		\begin{eqnarray}\label{eq:operators1}
			\ket{0 , \Lambda_\eta^\epsilon} = \lim_{T \rightarrow \infty} \frac{e^{-h T} O_{\Lambda_\eta^\epsilon} | \Omega \rangle}{\Big|e^{-h T} O_{\Lambda_\eta^\epsilon} | \Omega \rangle\Big|}
		\end{eqnarray}
		with $h$ denoting the Hamiltonian,
		\begin{eqnarray}\label{eq:operators2}
			O_{\Lambda_\eta^\epsilon} = \bar{Q}(-r/2) U(-r/2;0) \mathcal{B}_{\Lambda_\eta^\epsilon}(0) U(0;+r/2) Q(+r/2)
		\end{eqnarray}
		($Q$ and $\bar{Q}$ are static quark and static antiquark operators and $U$ represents a straight parallel transporter)
		and
		\begin{eqnarray}\label{eq:operators3}
			\mathcal{B}_{\Sigma_g^+} = 1 \quad , \quad \mathcal{B}_{\Sigma_u^-} = -i B_z \quad , \quad \mathcal{B}_{\Pi_u^+} = -i B_x \quad , \quad \mathcal{B}_{\Pi_u^-} = +i B_y
		\end{eqnarray}
		($B_j=- \epsilon_{ijk}{F}_{jk}/2$ is the chromomagnetic field operator).
		The $x$ and $y$ coordinates of quark, antiquark and chromomagnetic field positions are $0$ and have been omitted for better readability, i.e.\ $\pm r/2 \equiv (0,0,\pm r/2)$.
		$c_F$ is a necessary matching coefficient for the chromomagnetic fields originating from NRQCD \cite{Eichten:1990vp,Brambilla:2023vwm}.

		Note that Eqs.\ \eqref{eq:Vsa} to \eqref{eq:VmixSigma} depend on the definition of the operators $\mathcal{B}_{\Lambda_\eta^\epsilon}$, because of Eqs.\ \eqref{eq:operators1} and \eqref{eq:operators2}. A choice of operators $\mathcal{B}_{\Lambda_\eta^\epsilon}$ different from Eq.\ \eqref{eq:operators3}, but still generating states with definite quantum numbers $\Lambda_\eta^\epsilon$, might lead to different signs or even phases in Eqs.\ \eqref{eq:Vsa} to \eqref{eq:VmixSigma}. This will be relevant in Section~\ref{sec:latticeoperators}, where we discuss suitable lattice creation operators.
		
		
		\subsection{Definition of lattice creation operators}\label{sec:latticeoperators}

		The creation operators defined in Eqs.\ \eqref{eq:operators2} and \eqref{eq:operators3} are not practical for lattice computations of hybrid spin-dependent potentials and hybrid-quarkonium mixing potentials, i.e.\ for computations of the matrix elements on the right hand sides of Eqs.\ \eqref{eq:Vsa} to \eqref{eq:VmixSigma},
		due to the poor ground state overlaps they generate.
		We rather employ optimized lattice creation operators from Ref.\ \cite{Capitani:2018rox}, which allow to extract the potentials from generalized Wilson loops with smaller temporal extents, where the noise-to-signal ratio is favorable. For $\Lambda_\eta^\epsilon = \Sigma_g^+ , \Sigma_u^- , \Pi_u^+$ these operators are of the form
		\begin{eqnarray}
			O_{\Lambda_\eta^\epsilon}^\text{lattice} = \bar{Q}(-r/2) a_{S;\Lambda_\eta^\epsilon}(-r/2,+r/2) Q(+r/2) , \label{eq:lattice_operator}
		\end{eqnarray}
		where $a_{S;\Lambda_{\eta}^{\epsilon}}$ is given by a sum of properly transformed spatial insertions $ S(r_1,r_2)$,
		\begin{eqnarray}
			\nonumber & & \hspace{-0.7cm} a_{S; \Lambda_{\eta}^{\epsilon}} (-r/2, +r/2) = 
			\frac{1}{4} \Big(1 + \eta (\mathcal{P} \circ \mathcal{C}) + \epsilon \mathcal{P}_x + \eta \epsilon (\mathcal{P} \circ \mathcal{C}) \mathcal{P}_x\Big) \\
			\nonumber & & \hspace{0.7cm} \sum_{k=0}^{3} \frac{1}{2} \left(\exp(\frac{+i\pi \Lambda k}{2}) + \exp(\frac{-i\pi \Lambda k}{2})\right) R\left( \frac{\pi k}{2}\right)  
	    \Big(U(-r/2,r_1) S(r_1,r_2)  U(r_2,+r/2)\Big) . \\
		   & & \label{eq:new_def_lattice_creationoperator} 
		\end{eqnarray}
		The insertions $S$ were designed and optimized in Ref.\ \cite{Capitani:2018rox} and employed for the lattice computation of hybrid static potentials in Refs.\ \cite{Capitani:2018rox,Schlosser:2021wnr}.
		We follow Ref. \cite{Schlosser:2021wnr} and use $S = S_\text{\RN{4},2}$ for $\Lambda_\eta^\epsilon = \Sigma_u^-$ and $S = S_\text{\RN{3},1}$ for $\Lambda_\eta^\epsilon = \Pi_u^+$ as defined in detail in Ref.\ \cite{Capitani:2018rox}.
		For $\Pi_u^-$, we use for $S$ a counterclockwise $\pi/2$ rotation of the insertion $S_\text{\RN{3},1}$ with respect to the $z$ axis.
		Consequently, the operators $O_{\Pi_u^+}^\text{lattice}$ and $O_{\Pi_u^-}^\text{lattice}$ have the same structure and are related by a $\pi/2$ rotation.
		We note that $O_{\Pi_u^-}^\text{lattice}$ is different from an operator obtained via Eqs.\ (\ref{eq:lattice_operator}) and (\ref{eq:new_def_lattice_creationoperator}) for $\Pi_u^-$ using the non-rotated insertion $S_\text{\RN{3},1}$.
		This alternative operator, which we denote as $O_{\Pi_u^+}^{\text{lattice},2}$, is also valid, i.e.\ generates states with quantum numbers $\Pi_u^-$, but is numerically less efficient.
		We used $O_{\Pi_u^+}^{\text{lattice},2}$ only for numerical checks of our code and results.
		For $\Lambda_\eta^\epsilon = \Sigma_g^+$, $a_{S;\Sigma_g^+}$ is just a straight path of gauge links.

		As investigated and discussed in Section~4.2 of Ref.\ \cite{Capitani:2018rox}, link smearing drastically increases the ground state overlaps generated by the operators (\ref{eq:lattice_operator}). We use APE smearing as defined in Section~3.1.3 of Ref.\ \cite{Jansen:2008si} with parameters $\alpha_\text{APE} = 0.5$ and $N_\text{APE} = 50$. These are exactly the same parameters we used in Ref.\ \cite{Schlosser:2021wnr} for the computation of the ordinary static potential $V_{\Sigma_g^+}(r)$ and the hybrid static potentials $V_{\Sigma_u^-}(r)$ and $V_{\Pi_u}(r)$ on the gauge link ensemble with $a = 0.060 \, \textrm{fm}$ discussed in Section~\ref{SEC_ensemble}. 
		
		To be able to use Eqs.\ \eqref{eq:Vsa} to \eqref{eq:VmixSigma}, it is mandatory that the lattice operators $O_{\Lambda_\eta^\epsilon}^\text{lattice}$ are consistent with Eqs.\ \eqref{eq:operators1} to \eqref{eq:operators3}, i.e.\ they have to satisfy the condition
		\begin{eqnarray}\label{eq:groundstate}
			\ket{0 , \Lambda_\eta^\epsilon} = \lim_{T \rightarrow \infty} \frac{e^{-h T} O_{\Lambda_\eta^\epsilon}^\text{lattice} | \Omega \rangle}{\Big|e^{-h T} O_{\Lambda_\eta^\epsilon}^\text{lattice} | \Omega \rangle\Big|} ,
		\end{eqnarray}
		where $| 0 , \Lambda_\eta^\epsilon \rangle$ is mathematically identical to the the ground state $| 0 , \Lambda_\eta^\epsilon \rangle$ appearing on the left hand side of Eq.\ \eqref{eq:operators1} (physical equivalence where phases can be different is not sufficient).
		The consistency between the operator definitions can be guaranteed by the replacement
		\begin{eqnarray}\label{eq:phasealignment_operator}
			O_{\Lambda_\eta^\epsilon}^\text{lattice} \ \ \rightarrow \ \ \underbrace{\Bigg(\lim_{T \rightarrow \infty} \frac{\langle \Omega | (O_{\Lambda_\eta^\epsilon}^\text{lattice})^\dagger e^{-h T} O_{\Lambda_\eta^\epsilon} | \Omega \rangle}{\Big|\langle \Omega | (O_{\Lambda_\eta^\epsilon}^\text{lattice})^\dagger e^{-h T} O_{\Lambda_\eta^\epsilon} | \Omega \rangle\Big|}\Bigg)}_{= \alpha_{\Lambda_\eta^\epsilon}} O_{\Lambda_\eta^\epsilon}^\text{lattice} .
		\end{eqnarray}
		It is straightforward to show that $\langle \Omega | (O_{\Lambda_\eta^\epsilon}^\text{lattice})^\dagger e^{-h T} O_{\Lambda_\eta^\epsilon} | \Omega \rangle$ is real. 
		Consequently, the unknown phases $\alpha_{\Lambda_\eta^\epsilon}$ can only take values $+1$ or $-1$ for the operators we use. 
		A simple and quick lattice computation revealed $\alpha_{\Sigma_g^+} = \alpha_{\Sigma_u^-} = \alpha_{\Pi_u^+} = \alpha_{\Pi_u^-} = +1$.


		\subsection{Computation of matrix elements}\label{sec:computation_matrixelements}
		
		Eqs.\ \eqref{eq:Vsa} to \eqref{eq:VmixSigma} relate the hybrid spin-dependent potentials and the hybrid-quarkonium mixing potentials to matrix elements in SU(3) gauge theory. We extract these matrix elements from ratios of Wilson loop-like correlation functions defined as
		\begin{eqnarray}
			\nonumber & & \hspace{-0.7cm} R^{B_k}_{\Lambda_\eta^\epsilon{\Lambda_\eta^\epsilon}'}(t;r,T) \\
			\label{eq:ratioRBk} & &  =W^{B_k}_{\Lambda_\eta^\epsilon{\Lambda_\eta^\epsilon}'}(t;r,T) 
			\bigg(\frac{1}{W_{ \Lambda_{\eta}^\epsilon}(r,T)W_{ {\Lambda_\eta^\epsilon}'}(r,T)}\bigg)^{1/2}
			\bigg(\frac{W_{ {\Lambda_\eta^\epsilon}'}(r,T/2-t)W_{ \Lambda_{\eta}^\epsilon}(r,T/2+t) }{W_{ {\Lambda_{\eta}^\epsilon}}(r,T/2-t)W_{ {\Lambda_\eta^\epsilon}'}(r,T/2+t)}\bigg)^{1/2} .
		\end{eqnarray}
		$W_{\Lambda_\eta^\epsilon}(r,T)$ denotes the familiar (hybrid) Wilson loop, i.e.\ a correlation function of (hybrid) trial states with quantum numbers $\Lambda_{\eta}^\epsilon$,
		\begin{eqnarray}\label{eq:CorrelationfunctionW}
			\nonumber & & \hspace{-0.7cm} W_{\Lambda_{\eta}^\epsilon}(r,T) = \langle \Omega | (O_{\Lambda_\eta^\epsilon}^\text{lattice})^\dagger (r,T/2) O_{\Lambda_\eta^\epsilon}^\text{lattice} (r,-T/2) | \Omega \rangle  \\
			\nonumber & & = \Bigl\langle
				\Tr\Big(
				a_{S;\Lambda_{\eta}^{\epsilon}}(-r/2,+r/2;-T/2) U(+r/2;-T/2,T/2) \left(a_{S;\Lambda_{\eta}^{\epsilon}}(-r/2,+r/2;T/2)\right)^{\dagger} \\
				 & & \hspace{0.7cm} U(-r/2;T/2,-T/2)
				\Big)
			\Bigr\rangle_U 
			,
		\end{eqnarray}
		where $U(r;t_1,t_2)$ is a straight path of temporal gauge links from time $t_1$ to time $t_2$ at spatial position $\mathbf{r} = (0,0,r)$ and $\expval{\ldots}_U$ denotes the average on the ensemble of gauge link configurations discussed in Section~\ref{SEC_ensemble}.		
		$W^{B_k}_{\Lambda_\eta^\epsilon{\Lambda_\eta^\epsilon}'}(t;r,T)$ denotes a generalized Wilson loop containing two different (hybrid) creation operators with a chromomagnetic field insertion along one of its temporal lines,
		\begin{eqnarray}\label{eq:Wilsonloopcorrelator_with_insertion}
			\nonumber & & \hspace{-0.7cm} W^{B_k}_{ \Lambda_{\eta}^{\epsilon}{\Lambda_{\eta}^{\epsilon}}'}(t;r,T) = \langle \Omega | (O_{\Lambda_\eta^\epsilon}^\text{lattice})^\dagger (r,T/2) B^\text{lattice}_k(-{r}/2,t) O_{{\Lambda_\eta^\epsilon}^\prime}^\text{lattice} (r,-T/2) | \Omega \rangle  \\
			\nonumber & & = \Bigl\langle \Tr \Big(a_{S',{\Lambda_{\eta}^{\epsilon}}'}(-r/2,r/2; -T/2) U(r/2; -T/2,T/2) \left(a_{S,{\Lambda_{\eta}^{\epsilon}}}(-r/2,r/2; T/2)\right)^\dagger \\
			 & & \hspace{0.7cm} U(-r/2; T/2,t)B^\text{lattice}_k(-{r}/2,t)U(-r/2; t,-T/2))\Big)\Bigr\rangle_U 
			 .
		\end{eqnarray}
		The lattice chromomagnetic field $B^\text{lattice}_k$ with $k=x,y,z$ is always inserted into the temporal line at $-r/2 \equiv (0,0,-r/2)$ with $t$ denoting its temporal position (in the following we omit the arguments $(-r/2,t)$ to keep equations short and clear).
		We note that generalized Wilson loops with a chromomagnetic field inserted at $-r/2$ and $+r/2$ are related by a $\mathcal{P}\circ \mathcal{C}$ transformation.
		
		We use the cloverleaf discretization for a chromomagnetic field insertion on the lattice, i.e.\
		\begin{equation}\label{eq:def_discretization_Bfield}
			{B}^\text{lattice}_l = -\frac{\epsilon_{lmn}}{2}\left(\Pi_{mn}-\Pi_{mn}^\dagger\right) = -igB_l
		\end{equation}
		with the clover-leaf plaquette
		\begin{equation}
			\label{eq:def_discretization_Bfield_2} \Pi_{mn} = \frac{1}{4}\Big(P_{m,n}+P_{n,-m}+P_{-m,-n}+ P_{-n,m} \Big) .
		\end{equation}
		
		The Wilson loops are symmetrically placed around $t=0$.
		Thus, for odd $T$ the center of the temporal line at $t=0$ is located between two lattice sites.
		Since the chromomagnetic field insertion must be placed at a lattice site, integer $t/a= \dots,-1,\,  0,\,1,\dots$ are only possible for even $T$, while half-integer $t/a=\dots,-3/2,\, -1/2,\,1/2,\,3/2,\dots$ are only possible for odd $T$.
		$t=0$ corresponds to a symmetric insertion between the two spatial transporters at $-T/2$ and $T/2$ of the Wilson loop.
		Thus, for $t< 0$ the temporal distance to the spatial transporter at $-T/2$ is smaller than to the spatial transporter at $+T/2$. For $t>0$ the situation is reversed.
		
		By inserting the spectral decompositions of the correlators into Eq.\ \eqref{eq:ratioRBk} one can show that the correlator ratio $R^{B_k}_{\Lambda_\eta^\epsilon{\Lambda_\eta^\epsilon}'}(t;r,T)$ yields the matrix elements of interest in the large $T$ limit.
		These spectral decompositions are given by
		\begin{eqnarray}
			\nonumber & & \hspace{-0.7cm} W^{B_k}_{ \Lambda_\eta^\epsilon{\Lambda_\eta^\epsilon}'}(t;r,T) = 
				\sum_n \sum_m \langle \Omega | (O_{\Lambda_\eta^\epsilon}^\text{lattice})^\dagger\ket{n,{\Lambda_\eta^\epsilon}}\bra{n,{\Lambda_\eta^\epsilon}}B^\text{lattice}_k\ket{m,{{\Lambda_\eta^\epsilon}'}}(r)\bra{m,{{\Lambda_\eta^\epsilon}'}} O_{{\Lambda_\eta^\epsilon}^\prime}^\text{lattice} | \Omega \rangle \nonumber \\
			\label{EQN_specWB} & & \hspace{0.7cm} e^{-(V_{n,{\Lambda_\eta^\epsilon}}(r) + V_{m,{\Lambda_\eta^\epsilon}'}(r))T/2} e^{+( V_{n,{\Lambda_\eta^\epsilon}}(r) - V_{m,{\Lambda_\eta^\epsilon}'}(r))t} \\
			\label{EQN_specW} & & \hspace{-0.7cm} W_{ \Lambda_\eta^\epsilon}(r,T) =	
			\sum_n \langle \Omega | (O_{\Lambda_\eta^\epsilon}^\text{lattice})^\dagger\ket{n,{\Lambda_\eta^\epsilon}} \bra{n,{\Lambda_\eta^\epsilon}} O_{{\Lambda_\eta^\epsilon}}^\text{lattice} | \Omega \rangle e^{-V_{n,{\Lambda_\eta^\epsilon}}(r) T}
		\end{eqnarray}
		with $\bra{n,\Lambda_\eta^\epsilon}B^\text{lattice}_k\ket{m,{\Lambda_\eta^\epsilon}'} (r) 
		= -ig \bra{n,\Lambda_\eta^\epsilon}B_k\ket{m,{\Lambda_\eta^\epsilon}'} (r)$.
		In the large $T$ limit, excited states are suppressed and the ground states dominate.
		The first square root in Eq.\ \eqref{eq:ratioRBk} compensates the overlaps $\langle \Omega | (O_{\Lambda_\eta^\epsilon}^\text{lattice})^\dagger\ket{0,{\Lambda_\eta^\epsilon}}$ and $\bra{0,{{\Lambda_\eta^\epsilon}'}}O_{{\Lambda_\eta^\epsilon}^\prime}^\text{lattice} | \Omega \rangle$ and the exponential $T$-dependence of the ground state contribution.
		The second square root in Eq.\ \eqref{eq:ratioRBk} cancels the exponential $t$-dependence.
		In the large $T$ limit the correlator ratio \eqref{eq:ratioRBk}, thus, approaches one of the matrix elements,
		\begin{align}
			\label{EQN_R_matrix_element}
			\lim_{T\to \infty}R^{B_k}_{\Lambda_\eta^\epsilon{\Lambda_\eta^\epsilon}'}(t;r,T)  =& \bra{0,\Lambda_\eta^\epsilon}B^\text{lattice}_k\ket{0,{\Lambda_\eta^\epsilon}'} (r) 
			= -ig \bra{0,\Lambda_\eta^\epsilon}B_k\ket{0,{\Lambda_\eta^\epsilon}'} (r) .
		\end{align}
		Note that the matrix elements depend on the spatial separation $r$ of the static quark and antiquark, but are independent of the temporal position $t$ of the chromomagnetic field insertion. It is, however, useful to have that $t$-dependence in the correlator ratio $R^{B_k}_{\Lambda_\eta^\epsilon{\Lambda_\eta^\epsilon}'}(t;r,T)$, to generate a larger set of data points converging to $\bra{0,\Lambda_\eta^\epsilon}B^\text{lattice}_k\ket{0,{\Lambda_\eta^\epsilon}'} (r)$. This allows to check more rigorously, whether excited states are indeed negligible. Having more data points for $R^{B_k}_{\Lambda_\eta^\epsilon{\Lambda_\eta^\epsilon}'}(t;r,T)$ might also reduce statistical uncertainties in the final results.
		
		
		\subsection{Gradient flow}\label{sec:gradientflow}

		The bare chromomagnetic matrix elements $\bra{0,\Lambda_\eta^\epsilon} B_k \ket{0,{\Lambda_\eta^\epsilon}'}(r)$ appearing in Eqs.\ (\ref{eq:Vsa}) to (\ref{eq:VmixSigma}) are infinite and need to be renormalized. This is done by multiplication with a matching coefficient $c_F$.
		
		One possibility of renormalization suited for lattice gauge theory computations is to use gradient flow \cite{Luscher:2010iy}.
		One first has to compute a matrix element $\bra{0,\Lambda_\eta^\epsilon} B_k \ket{0,{\Lambda_\eta^\epsilon}'}(r)$ at several lattice spacings $a > 0$ and flow times $t_f > 0$.
		By multiplication with an appropriate matching coefficient $c_F(t_f,\mu)$ these data points can be converted from the gradient flow scheme at flow time $t_f$ to the $\overline{\text{MS}}$ scheme at scale $\mu$ \footnote{The matching coefficients $c_F(t_f,\mu)$ are known up to one-loop order \cite{Brambilla:2023vwm,Altenkort:2024spl}.}.
		Then one can carry out a continuum extrapolation of $c_f(t_f,\mu) \bra{0,\Lambda_\eta^\epsilon} B_k \ket{0,{\Lambda_\eta^\epsilon}'}(r)$ at fixed flow time $t_f$, which gives a finite result. This step has to be repeated for several flow times $t_f$. Finally, these continuum results are extrapolated to flow time $t_f = 0$. The result of this extrapolation is the renormalized chromomagnetic matrix element in the $\overline{\text{MS}}$ scheme at scale $\mu$, which is equivalent to one of the $(1/m_Q)^1$ potentials at separation $r$, as expressed by Eqs.\ (\ref{eq:Vsa}) to (\ref{eq:VmixSigma}).
		
		Note, however, that we do not perform such a renormalization procedure in this work. 
		Our aim is rather to carry out a first exploratory computation of chromomagnetic matrix elements at a single lattice spacing $a = 0.060 \, \text{fm}$ and to demonstrate that, within our chosen setup, we are able to reach sufficient statistical precision for a future combined continuum and zero-flow time extrapolation.
		
		Another advantage of using gradient flow is the drastic reduction of statistical errors. The reason for that is, that gradient flow can be interpreted as a smearing procedure with radius $r_f = \sqrt{8 t_f}$, also referred to as flow radius.
		For small Wilson loops, however, gradient flow might introduce sizable systematic errors, because of overlapping smeared gauge links from opposite spatial or temporal lines. For ordinary Wilson loops without chromomagnetic field insertions, $W_{\Lambda_\eta^\epsilon}(r,T)$, these systematic errors are expected to be mild, if $r,T \gtapprox 2 r_f$.
		For generalized Wilson loops with a chromomagnetic field insertion, $W^{B_k}_{ \Lambda_\eta^\epsilon{\Lambda_\eta^\epsilon}'}(t;r,T)$, the constraint on $T$ is more restrictive, $T/2 - \abs{t} \gtapprox 2 r_f$.
		Furthermore, chromomagnetic field insertions $B_x$ and $B_y$, discretized by cloverleafs, extend by $\pm a$ along the axis of separation (the $z$ axis), which leads to the more severe constraint $r \gtapprox 2 r_f + a$.
		
		We note that gradient flow has been used in the context of related correlators in Refs.\ \cite{Altenkort:2019qnr,Leino:2021bpz,Leino:2021vop,Eichberg:2023trq,Altenkort:2024spl,Eichberg:2024RL}.


	\newpage
		
		
	\section{\label{sec:results}Numerical results}
	
	
		\subsection{Hybrid spin-dependent potentials and hybrid-quarkonium mixing potentials at flow radius $r_f/a = 1.8$}\label{sec:extraction_matrixelements}

		We computed the correlator ratios $R^{B_k}_{\Lambda_\eta^\epsilon{\Lambda_\eta^\epsilon}'}(t;r,T)$ defined in Eq.\ (\ref{eq:ratioRBk}) on the ensemble of gauge link configurations with lattice spacing $a = 0.060 \, \text{fm}$ discussed in Section~\ref{SEC_ensemble}.
		To enhance statistical precision, both translational and rotational symmetries were exploited, including appropriate averaging of correlator ratios involving the operators $O_{\Pi_u^+}^\text{lattice}$ and $O_{\Pi_u^-}^\text{lattice}$, as expressed by Eqs.\ (\ref{eq:Vsb}) and (\ref{eq:VmixPi}) for the corresponding matrix elements.

\begin{figure}[p]
			\begin{center}
			\vspace{0.1cm}
			{\small \textbf{correlator ratios for hybrid spin-dependent potentials}}

			\vspace{0.1cm}
			\includegraphics[width=0.48\linewidth,page=4]{Figure1_tmin_eq_max_trf_9}
			\includegraphics[width=0.48\linewidth,page=3]{Figure1_tmin_eq_max_trf_9}

			\vspace{0.1cm}
			{\small \textbf{correlator ratios for hybrid-quarkonium mixing potentials}}

			\vspace{0.1cm}
			\includegraphics[width=0.48\linewidth,page=1]{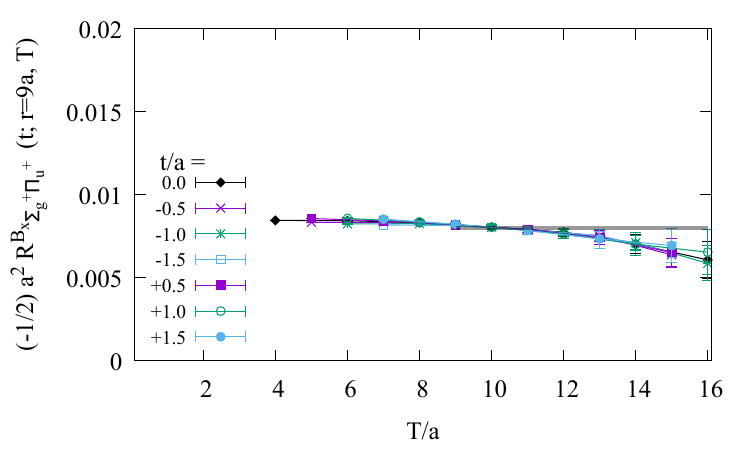}
			\includegraphics[width=0.48\linewidth,page=2]{Figure1_tmin_eq_max_trf_9}
			\end{center}
			\caption{Correlator ratios
			$-R^{B_z}_{\Pi_u^- \Pi_u^+}(t;r=9a,T)$, $-R^{B_y}_{\Sigma_u^- \Pi_u^+}(t;r=9a,T)$,
			$-(1/2) R^{B_x}_{\Sigma_g^+ \Pi_u^+}(t;r=9a,T)$ and $-(1/2) R^{B_z}_{\Sigma_g^+ \Sigma_u^-}(t;r=9a,T)$
			in units of the lattice spacing $a = 0.060 \, \text{fm}$  at flow radius $r_f / a = 1.8$. The prefactors $-1$ and $-1/2$ were chosen such that the asymptotic values at large $T$ correspond to $V^{sa}_{11}(r=9a) / c_F$ (top left), $V^{sb}_{10}(r=9a) / c_F$ (top right), $V^\text{mix}_{\Sigma_u^-}(r=9a) m_Q / c_F$ (bottom left) and $V^\text{mix}_{\Pi_u}(r=9a) m_Q / c_F$ (bottom right), see Eqs.\ (\ref{EQN_R_matrix_element}) and (\ref{eq:Vsa}) to (\ref{eq:VmixSigma}). Each grey band represents a fit of a constant to data points fulfilling both $T/2 - \abs{t} \ge 2 r_f$ and $9 a \leq T \le T_\text{max}$.}
			\label{fig:ratios_effectivepotentials1}
		\end{figure}

		In this subsection we present and discuss results obtained at flow radius $r_f / a = 1.8$.
		In Figure~\ref{fig:ratios_effectivepotentials1} we show the ratios $R^{B_k}_{\Lambda_\eta^\epsilon{\Lambda_\eta^\epsilon}'}(t;r,T)$ as functions of $T$ for the exemplary spatial separation $r = 9a = 0.54 \, \text{fm}$ and several temporal positions of the chromomagnetic field $t/a = -3/2, \, -1, \, -1/2, \, 0, \, +1/2, \, +1, \, +3/2$.
		To cross-check our numerical results, we also computed ratios using the operator $O_{\Pi_u^+}^{\text{lattice},2}$ (these ratios are not shown) instead of $O_{\Pi_u^+}^\text{lattice}$ and verified that the obtained asymptotic values at large $T$ are compatible within statistical errors.

		The ratios related to hybrid spin-dependent potentials (top row) exhibit a slight negative slope at small $T$ and seem to reach their asymptotic values later than the ratios related to hybrid-quarkonium mixing potentials (bottom row).
		This indicates that the creation operator $O_{\Sigma_g^+}^\text{lattice}$ generates a larger ground state overlap than its hybrid counterparts $O_{\Sigma_u^-}^\text{lattice}$ and $O_{\Pi_u^\pm}^\text{lattice}$.
		This is not surprising, because the $\Sigma_g^+$ ground state has a rather simple cigar-shaped gluon distribution, which is approximated rather well by a straight path of APE-smeared links.
		The hybrid ground states, on the other hand, have more complicated structures as investigated in detail in Refs.\ \cite{Bicudo:2018jbb,Muller:2019joq}.
		
		To extract a matrix element, or equivalently one of the potentials $V^{sa}_{11}(r) / c_F$, $V^{sb}_{10}(r) / c_F$, \\ $V^\text{mix}_{\Sigma_u^-}(r) m_Q / c_F$ and $V^\text{mix}_{\Pi_u}(r) m_Q / c_F$ (see Eqs.\ (\ref{eq:Vsa}) to (\ref{eq:VmixSigma})), at a given separation $r$, we fit a constant to a set of ratios $R^{B_k}_{\Lambda_\eta^\epsilon{\Lambda_\eta^\epsilon}'}(t;r,T)$ comprising several $t$ and and sufficiently large $T$.
		To avoid unwanted effects from overlapping operators (including both the creation operators at time $\pm T/2$ and the chromomagnetic field at time $t$), which are smeared in time direction when applying gradient flow, we always exclude ratios where $T/2 - \abs{t} < 2 r_f$ (see the discussion in Section~\ref{sec:gradientflow}).
		To assure that the fit result corresponds to the asymptotic value of $R^{B_k}_{\Lambda_\eta^\epsilon{\Lambda_\eta^\epsilon}'}(t;r,T)$ at large $T$, we further restricted the ratios entering a fit.
		We checked the stability of our numerical results by exploring different strategies, which are discussed in the following.

		Following the method in our previous works \cite{Capitani:2018rox,Schlosser:2021wnr,Herr:2023xwg}, we consider only those of the remaining ratios $R^{B_k}_{\Lambda_\eta^\epsilon{\Lambda_\eta^\epsilon}'}(t;r,T)$ with $T_\text{min}(t,r) \leq T \le T_\text{max}$ and $T_\text{min}(t,r)$ and $T_\text{max}$ chosen as follows:
		\begin{itemize}
			\item $T_\text{min}(t,r)$ is the smallest $T$, where $|R^{B_k}_{\Lambda_\eta^\epsilon{\Lambda_\eta^\epsilon}'}(t;r,T) - R^{B_k}_{\Lambda_\eta^\epsilon{\Lambda_\eta^\epsilon}'}(t;r,T+a)| < 2 \Delta R^{B_k}_{\Lambda_\eta^\epsilon{\Lambda_\eta^\epsilon}'}(t;r,T+a)$ with $\Delta R^{B_k}_{\Lambda_\eta^\epsilon{\Lambda_\eta^\epsilon}'}(t;r,T+a)$ denoting the statistical error of $R^{B_k}_{\Lambda_\eta^\epsilon{\Lambda_\eta^\epsilon}'}(t;r,T+a)$.

			\item $T_\text{max}$ is the largest $T$, where statistical errors are still on a moderate level, i.e.\ the noise-to-signal ratio is significantly below $1$. In contrast to $T_\text{min}(t,r)$, $T_\text{max}$ has an almost negligible effect on the fit results. We chose $T_{\max} / a = 14$ for the ratios $R^{B_z}_{\Pi_u^- \Pi_u^+}(t;r,T)$ and $R^{B_y}_{\Sigma_u^- \Pi_u^+}(t;r,T)$ related to hybrid spin-dependent potentials and $T_{\max} / a = 16$ for the ratios $R^{B_x}_{\Sigma_g^+ \Pi_u^+}(t;r,T)$ and $R^{B_z}_{\Sigma_g^+ \Sigma_u^-}(t;r,T)$ related to hybrid-quarkonium mixing potentials.
		\end{itemize}
		For the majority of fits this method determines $T_\text{min}(t,r)$ as the smallest possible $T$ satisfying the constraint $T/2 - \abs{t} \ge 2 r_f$, where reasonable fits are indicated by reduced $\chi^2$ values of order $1$.
		In particular, $T_\text{min}(t=0,r) / a = 8$ for most fits.
		However, a drawback of this method is that the statistical errors of the extracted potential values increase significantly if $T_\text{min}(t,r)$ is determined to be larger.
		For instance, when extracting the potential $V^{sb}_{10}(r)$, the algorithm selects $T_\text{min}(t=0,r) / a = 10$ for $ 3 \leq r/a \leq 7$ and $T_\text{min}(t=0,r) / a = 8$ for other values of $r/a$.
		As a result, neighboring potential data points have errors of significantly different magnitude.
		This discrepancy is not due to a substantial difference in the data quality but rather arise from a sudden change in $T_\text{min}(t,r)$.
		Having such a strong variation in the errors is certainly not ideal, e.g.\ when fitting a parametrization to the potential data points.

		An alternative method avoiding this problem is to restrict the ratios $R^{B_k}_{\Lambda_\eta^\epsilon{\Lambda_\eta^\epsilon}'}(t;r,T)$ using a fixed $T_\text{min}$ for all $t$ and $r$, i.e.\ the ratios are restricted by  $T_\text{min} \leq T \le T_\text{max}$.
		As before, ratios are furthermore restricted by $T/2 - \abs{t} \ge 2r_f$.
		We extracted the potentials using $T_\text{min} / a = 8$, $T_\text{min} / a = 9$ and $T_\text{min} / a = 10$, respectively.
		$T_\text{min} / a = 8$ and $T_\text{min} / a = 9$ are typically suggested by the previously discussed algorithm.
		While the potentials extracted with these two choices agree within statistical uncertainties, there appears to be a slight trend that the potentials determined with $T_\text{min} / a = 9$  tend to be slightly lower than those obtained with $T_\text{min} / a = 8$.
		For $T_\text{min} / a = 10$, statistical errors increase further and results are again compatible with those corresponding to $T_\text{min} / a = 8$ and $T_\text{min} / a = 9$.
		Based on these observations, using the alternative strategy with $T_\text{min} / a = 9$ seems to be the best choice to extract the potentials including a fair and realistic estimate of statistical uncertainties.

		The numerical values for the potentials $V^{sa}_{11}(r) / c_F$, $V^{sb}_{10}(r) / c_F$, $V^\text{mix}_{\Sigma_u^-}(r) m_Q / c_F$ and $V^\text{mix}_{\Pi_u}(r) m_Q / c_F$ obtained with all four fitting approaches ($T_\text{min}(t,r)$ as discussed above, $T_\text{min} / a = 8$, $T_\text{min} / a = 9$ and $T_\text{min} / a = 10$) are collected in Table~\ref{TAB_numerical_results_Vsa_Vsb} and Table~\ref{TAB_numerical_results_Vmix}.


		\begin{table}[p]
		\def\arraystretch{1.2}
		\begin{center}
			\begin {tabular}{c|l|l|l|l}%
			\toprule \text {method} & \, $T_\text {min}(t,r)$ & $T_\text {min} / a = 8$ & $T_\text {min} / a = 9$ & $T_\text {min} / a = 10$ \\ \hline & \multicolumn {4}{c}{} \vspace {-0.5cm} \\ $r/a$ & \multicolumn {4}{c}{$a^2 V^{sa}_{11}(r) m_Q / c_F$} \\ & \multicolumn {4}{c}{} \vspace {-0.5cm} \\\midrule %
			\pgfutilensuremath {1}&\pgfmathprintnumber [fixed,fixed zerofill,precision=5]{1.7656387e-2}(\pgfmathprintnumber [fixed,fixed zerofill,precision=0]{5.7900009e1 })&\pgfmathprintnumber [fixed,fixed zerofill,precision=5]{1.7656387e-2}(\pgfmathprintnumber [fixed,fixed zerofill,precision=0]{5.7900009e1 })&\cellcolor {gray!50}\pgfmathprintnumber [fixed,fixed zerofill,precision=5]{1.7339691e-2}(\pgfmathprintnumber [fixed,fixed zerofill,precision=0]{9.748854e1 })&\pgfmathprintnumber [fixed,fixed zerofill,precision=5]{1.7278336e-2}(\pgfmathprintnumber [fixed,fixed zerofill,precision=0]{1.3135574e2 })\\%
			\pgfutilensuremath {2}&\pgfmathprintnumber [fixed,fixed zerofill,precision=5]{1.6526672e-2}(\pgfmathprintnumber [fixed,fixed zerofill,precision=0]{4.9174225e1 })&\pgfmathprintnumber [fixed,fixed zerofill,precision=5]{1.6526672e-2}(\pgfmathprintnumber [fixed,fixed zerofill,precision=0]{4.9174225e1 })&\cellcolor {gray!50}\pgfmathprintnumber [fixed,fixed zerofill,precision=5]{1.6242233e-2}(\pgfmathprintnumber [fixed,fixed zerofill,precision=0]{8.3333328e1 })&\pgfmathprintnumber [fixed,fixed zerofill,precision=5]{1.6310486e-2}(\pgfmathprintnumber [fixed,fixed zerofill,precision=0]{1.1806335e2 })\\%
			\pgfutilensuremath {3}&\pgfmathprintnumber [fixed,fixed zerofill,precision=5]{1.5048584e-2}(\pgfmathprintnumber [fixed,fixed zerofill,precision=0]{4.5881699e1 })&\pgfmathprintnumber [fixed,fixed zerofill,precision=5]{1.5048584e-2}(\pgfmathprintnumber [fixed,fixed zerofill,precision=0]{4.5881699e1 })&\cellcolor {gray!50}\pgfmathprintnumber [fixed,fixed zerofill,precision=5]{1.4695908e-2}(\pgfmathprintnumber [fixed,fixed zerofill,precision=0]{7.3156387e1 })&\pgfmathprintnumber [fixed,fixed zerofill,precision=5]{1.465451e-2}(\pgfmathprintnumber [fixed,fixed zerofill,precision=0]{1.0198471e2 })\\%
			\pgfutilensuremath {4}&\pgfmathprintnumber [fixed,fixed zerofill,precision=5]{1.3461426e-2}(\pgfmathprintnumber [fixed,fixed zerofill,precision=0]{4.3806976e1 })&\pgfmathprintnumber [fixed,fixed zerofill,precision=5]{1.3461426e-2}(\pgfmathprintnumber [fixed,fixed zerofill,precision=0]{4.3806976e1 })&\cellcolor {gray!50}\pgfmathprintnumber [fixed,fixed zerofill,precision=5]{1.2914734e-2}(\pgfmathprintnumber [fixed,fixed zerofill,precision=0]{6.8145905e1 })&\pgfmathprintnumber [fixed,fixed zerofill,precision=5]{1.2560562e-2}(\pgfmathprintnumber [fixed,fixed zerofill,precision=0]{9.4280884e1 })\\\hline %
			\pgfutilensuremath {5}&\pgfmathprintnumber [fixed,fixed zerofill,precision=5]{1.2048767e-2}(\pgfmathprintnumber [fixed,fixed zerofill,precision=0]{4.3117401e1 })&\pgfmathprintnumber [fixed,fixed zerofill,precision=5]{1.2048767e-2}(\pgfmathprintnumber [fixed,fixed zerofill,precision=0]{4.3117401e1 })&\cellcolor {gray!50}\pgfmathprintnumber [fixed,fixed zerofill,precision=5]{1.1474747e-2}(\pgfmathprintnumber [fixed,fixed zerofill,precision=0]{6.6433884e1 })&\pgfmathprintnumber [fixed,fixed zerofill,precision=5]{1.1041016e-2}(\pgfmathprintnumber [fixed,fixed zerofill,precision=0]{9.1102585e1 })\\%
			\pgfutilensuremath {6}&\pgfmathprintnumber [fixed,fixed zerofill,precision=5]{1.0929947e-2}(\pgfmathprintnumber [fixed,fixed zerofill,precision=0]{4.4159256e1 })&\pgfmathprintnumber [fixed,fixed zerofill,precision=5]{1.0929947e-2}(\pgfmathprintnumber [fixed,fixed zerofill,precision=0]{4.4159256e1 })&\cellcolor {gray!50}\pgfmathprintnumber [fixed,fixed zerofill,precision=5]{1.0597275e-2}(\pgfmathprintnumber [fixed,fixed zerofill,precision=0]{7.1054214e1 })&\pgfmathprintnumber [fixed,fixed zerofill,precision=5]{1.044136e-2}(\pgfmathprintnumber [fixed,fixed zerofill,precision=0]{1.0109222e2 })\\%
			\pgfutilensuremath {7}&\pgfmathprintnumber [fixed,fixed zerofill,precision=5]{9.7033173e-3}(\pgfmathprintnumber [fixed,fixed zerofill,precision=0]{4.8206512e1 })&\pgfmathprintnumber [fixed,fixed zerofill,precision=5]{9.7033173e-3}(\pgfmathprintnumber [fixed,fixed zerofill,precision=0]{4.8206512e1 })&\cellcolor {gray!50}\pgfmathprintnumber [fixed,fixed zerofill,precision=5]{9.4543915e-3}(\pgfmathprintnumber [fixed,fixed zerofill,precision=0]{8.2288345e1 })&\pgfmathprintnumber [fixed,fixed zerofill,precision=5]{9.4132263e-3}(\pgfmathprintnumber [fixed,fixed zerofill,precision=0]{1.1744873e2 })\\%
			\pgfutilensuremath {8}&\pgfmathprintnumber [fixed,fixed zerofill,precision=5]{8.344577e-3}(\pgfmathprintnumber [fixed,fixed zerofill,precision=0]{4.8867706e1 })&\pgfmathprintnumber [fixed,fixed zerofill,precision=5]{8.344577e-3}(\pgfmathprintnumber [fixed,fixed zerofill,precision=0]{4.8867706e1 })&\cellcolor {gray!50}\pgfmathprintnumber [fixed,fixed zerofill,precision=5]{7.9431e-3}(\pgfmathprintnumber [fixed,fixed zerofill,precision=0]{7.9876266e1 })&\pgfmathprintnumber [fixed,fixed zerofill,precision=5]{7.7307465e-3}(\pgfmathprintnumber [fixed,fixed zerofill,precision=0]{1.0614304e2 })\\%
			\pgfutilensuremath {9}&\pgfmathprintnumber [fixed,fixed zerofill,precision=5]{7.1638702e-3}(\pgfmathprintnumber [fixed,fixed zerofill,precision=0]{5.0399063e1 })&\pgfmathprintnumber [fixed,fixed zerofill,precision=5]{7.1638702e-3}(\pgfmathprintnumber [fixed,fixed zerofill,precision=0]{5.0399063e1 })&\cellcolor {gray!50}\pgfmathprintnumber [fixed,fixed zerofill,precision=5]{6.6021515e-3}(\pgfmathprintnumber [fixed,fixed zerofill,precision=0]{8.0605911e1 })&\pgfmathprintnumber [fixed,fixed zerofill,precision=5]{6.1413696e-3}(\pgfmathprintnumber [fixed,fixed zerofill,precision=0]{1.0699997e2 })\\%
			\pgfutilensuremath {10}&\pgfmathprintnumber [fixed,fixed zerofill,precision=5]{6.3766953e-3}(\pgfmathprintnumber [fixed,fixed zerofill,precision=0]{6.3429703e1 })&\pgfmathprintnumber [fixed,fixed zerofill,precision=5]{6.3766953e-3}(\pgfmathprintnumber [fixed,fixed zerofill,precision=0]{6.3429703e1 })&\cellcolor {gray!50}\pgfmathprintnumber [fixed,fixed zerofill,precision=5]{5.8975388e-3}(\pgfmathprintnumber [fixed,fixed zerofill,precision=0]{1.0438263e2 })&\pgfmathprintnumber [fixed,fixed zerofill,precision=5]{5.4991287e-3}(\pgfmathprintnumber [fixed,fixed zerofill,precision=0]{1.3758621e2 })\\%
			\pgfutilensuremath {11}&\pgfmathprintnumber [fixed,fixed zerofill,precision=5]{5.8858932e-3}(\pgfmathprintnumber [fixed,fixed zerofill,precision=0]{8.4151672e1 })&\pgfmathprintnumber [fixed,fixed zerofill,precision=5]{5.8858932e-3}(\pgfmathprintnumber [fixed,fixed zerofill,precision=0]{8.4151672e1 })&\cellcolor {gray!50}\pgfmathprintnumber [fixed,fixed zerofill,precision=5]{5.7985123e-3}(\pgfmathprintnumber [fixed,fixed zerofill,precision=0]{1.3804047e2 })&\pgfmathprintnumber [fixed,fixed zerofill,precision=5]{5.8267014e-3}(\pgfmathprintnumber [fixed,fixed zerofill,precision=0]{1.8250397e2 })\\%
			\pgfutilensuremath {12}&\pgfmathprintnumber [fixed,fixed zerofill,precision=5]{5.1228226e-3}(\pgfmathprintnumber [fixed,fixed zerofill,precision=0]{9.0807663e1 })&\pgfmathprintnumber [fixed,fixed zerofill,precision=5]{5.1228226e-3}(\pgfmathprintnumber [fixed,fixed zerofill,precision=0]{9.0807663e1 })&\cellcolor {gray!50}\pgfmathprintnumber [fixed,fixed zerofill,precision=5]{5.1196396e-3}(\pgfmathprintnumber [fixed,fixed zerofill,precision=0]{1.3802917e2 })&\pgfmathprintnumber [fixed,fixed zerofill,precision=5]{5.2721802e-3}(\pgfmathprintnumber [fixed,fixed zerofill,precision=0]{1.7701416e2 })\\\bottomrule %
			\end {tabular}%

			\vspace{1.0cm}
			
			\begin {tabular}{c|l|l|l|l}%
			\toprule \text {method} & \, $T_\text {min}(t,r)$ & $T_\text {min} / a = 8$ & $T_\text {min} / a = 9$ & $T_\text {min} / a = 10$ \\ \hline & \multicolumn {4}{c}{} \vspace {-0.5cm} \\ $r/a$ & \multicolumn {4}{c}{$a^2 V^{sb}_{10}(r) / c_F$} \\ & \multicolumn {4}{c}{} \vspace {-0.5cm} \\\midrule %
			\pgfutilensuremath {1}&\pgfmathprintnumber [fixed,fixed zerofill,precision=5]{1.7263367e-2}(\pgfmathprintnumber [fixed,fixed zerofill,precision=0]{4.2621704e1 })&\pgfmathprintnumber [fixed,fixed zerofill,precision=5]{1.7263367e-2}(\pgfmathprintnumber [fixed,fixed zerofill,precision=0]{4.2621704e1 })&\cellcolor {gray!50}\pgfmathprintnumber [fixed,fixed zerofill,precision=5]{1.6567673e-2}(\pgfmathprintnumber [fixed,fixed zerofill,precision=0]{7.69095e1 })&\pgfmathprintnumber [fixed,fixed zerofill,precision=5]{1.5963196e-2}(\pgfmathprintnumber [fixed,fixed zerofill,precision=0]{1.1853485e2 })\\%
			\pgfutilensuremath {2}&\pgfmathprintnumber [fixed,fixed zerofill,precision=5]{1.6217957e-2}(\pgfmathprintnumber [fixed,fixed zerofill,precision=0]{4.3467499e1 })&\pgfmathprintnumber [fixed,fixed zerofill,precision=5]{1.6217957e-2}(\pgfmathprintnumber [fixed,fixed zerofill,precision=0]{4.3467499e1 })&\cellcolor {gray!50}\pgfmathprintnumber [fixed,fixed zerofill,precision=5]{1.542958e-2}(\pgfmathprintnumber [fixed,fixed zerofill,precision=0]{7.3722305e1 })&\pgfmathprintnumber [fixed,fixed zerofill,precision=5]{1.4682175e-2}(\pgfmathprintnumber [fixed,fixed zerofill,precision=0]{1.1219025e2 })\\%
			\pgfutilensuremath {3}&\pgfmathprintnumber [fixed,fixed zerofill,precision=5]{1.3974777e-2}(\pgfmathprintnumber [fixed,fixed zerofill,precision=0]{5.7794083e1 })&\pgfmathprintnumber [fixed,fixed zerofill,precision=5]{1.4794891e-2}(\pgfmathprintnumber [fixed,fixed zerofill,precision=0]{3.5456818e1 })&\cellcolor {gray!50}\pgfmathprintnumber [fixed,fixed zerofill,precision=5]{1.3974777e-2}(\pgfmathprintnumber [fixed,fixed zerofill,precision=0]{5.7794083e1 })&\pgfmathprintnumber [fixed,fixed zerofill,precision=5]{1.3118423e-2}(\pgfmathprintnumber [fixed,fixed zerofill,precision=0]{8.7930893e1 })\\%
			\pgfutilensuremath {4}&\pgfmathprintnumber [fixed,fixed zerofill,precision=5]{1.2700226e-2}(\pgfmathprintnumber [fixed,fixed zerofill,precision=0]{4.5878601e1 })&\pgfmathprintnumber [fixed,fixed zerofill,precision=5]{1.3358246e-2}(\pgfmathprintnumber [fixed,fixed zerofill,precision=0]{2.8996902e1 })&\cellcolor {gray!50}\pgfmathprintnumber [fixed,fixed zerofill,precision=5]{1.2700226e-2}(\pgfmathprintnumber [fixed,fixed zerofill,precision=0]{4.5878601e1 })&\pgfmathprintnumber [fixed,fixed zerofill,precision=5]{1.2041855e-2}(\pgfmathprintnumber [fixed,fixed zerofill,precision=0]{6.9658081e1 })\\\hline %
			\pgfutilensuremath {5}&\pgfmathprintnumber [fixed,fixed zerofill,precision=5]{1.087001e-2}(\pgfmathprintnumber [fixed,fixed zerofill,precision=0]{5.4145355e1 })&\pgfmathprintnumber [fixed,fixed zerofill,precision=5]{1.1783936e-2}(\pgfmathprintnumber [fixed,fixed zerofill,precision=0]{3.0745255e1 })&\cellcolor {gray!50}\pgfmathprintnumber [fixed,fixed zerofill,precision=5]{1.1084686e-2}(\pgfmathprintnumber [fixed,fixed zerofill,precision=0]{4.7106537e1 })&\pgfmathprintnumber [fixed,fixed zerofill,precision=5]{1.0367203e-2}(\pgfmathprintnumber [fixed,fixed zerofill,precision=0]{6.956839e1 })\\%
			\pgfutilensuremath {6}&\pgfmathprintnumber [fixed,fixed zerofill,precision=5]{9.2052856e-3}(\pgfmathprintnumber [fixed,fixed zerofill,precision=0]{6.2216354e1 })&\pgfmathprintnumber [fixed,fixed zerofill,precision=5]{1.0182434e-2}(\pgfmathprintnumber [fixed,fixed zerofill,precision=0]{3.4631424e1 })&\cellcolor {gray!50}\pgfmathprintnumber [fixed,fixed zerofill,precision=5]{9.4660995e-3}(\pgfmathprintnumber [fixed,fixed zerofill,precision=0]{5.3386627e1 })&\pgfmathprintnumber [fixed,fixed zerofill,precision=5]{8.7087006e-3}(\pgfmathprintnumber [fixed,fixed zerofill,precision=0]{7.9570816e1 })\\%
			\pgfutilensuremath {7}&\pgfmathprintnumber [fixed,fixed zerofill,precision=5]{8.2882019e-3}(\pgfmathprintnumber [fixed,fixed zerofill,precision=0]{6.6233368e1 })&\pgfmathprintnumber [fixed,fixed zerofill,precision=5]{9.1539276e-3}(\pgfmathprintnumber [fixed,fixed zerofill,precision=0]{3.5240128e1 })&\cellcolor {gray!50}\pgfmathprintnumber [fixed,fixed zerofill,precision=5]{8.4850418e-3}(\pgfmathprintnumber [fixed,fixed zerofill,precision=0]{5.7338211e1 })&\pgfmathprintnumber [fixed,fixed zerofill,precision=5]{7.755426e-3}(\pgfmathprintnumber [fixed,fixed zerofill,precision=0]{8.7580948e1 })\\%
			\pgfutilensuremath {8}&\pgfmathprintnumber [fixed,fixed zerofill,precision=5]{8.2769653e-3}(\pgfmathprintnumber [fixed,fixed zerofill,precision=0]{3.8726746e1 })&\pgfmathprintnumber [fixed,fixed zerofill,precision=5]{8.2769653e-3}(\pgfmathprintnumber [fixed,fixed zerofill,precision=0]{3.8726746e1 })&\cellcolor {gray!50}\pgfmathprintnumber [fixed,fixed zerofill,precision=5]{7.8608765e-3}(\pgfmathprintnumber [fixed,fixed zerofill,precision=0]{6.3622833e1 })&\pgfmathprintnumber [fixed,fixed zerofill,precision=5]{7.5231369e-3}(\pgfmathprintnumber [fixed,fixed zerofill,precision=0]{9.1749222e1 })\\%
			\pgfutilensuremath {9}&\pgfmathprintnumber [fixed,fixed zerofill,precision=5]{7.2369339e-3}(\pgfmathprintnumber [fixed,fixed zerofill,precision=0]{4.1365494e1 })&\pgfmathprintnumber [fixed,fixed zerofill,precision=5]{7.2369339e-3}(\pgfmathprintnumber [fixed,fixed zerofill,precision=0]{4.1365494e1 })&\cellcolor {gray!50}\pgfmathprintnumber [fixed,fixed zerofill,precision=5]{7.077443e-3}(\pgfmathprintnumber [fixed,fixed zerofill,precision=0]{6.5604568e1 })&\pgfmathprintnumber [fixed,fixed zerofill,precision=5]{7.1115143e-3}(\pgfmathprintnumber [fixed,fixed zerofill,precision=0]{9.4700043e1 })\\%
			\pgfutilensuremath {10}&\pgfmathprintnumber [fixed,fixed zerofill,precision=5]{6.0621277e-3}(\pgfmathprintnumber [fixed,fixed zerofill,precision=0]{4.9577087e1 })&\pgfmathprintnumber [fixed,fixed zerofill,precision=5]{6.0621277e-3}(\pgfmathprintnumber [fixed,fixed zerofill,precision=0]{4.9577087e1 })&\cellcolor {gray!50}\pgfmathprintnumber [fixed,fixed zerofill,precision=5]{5.8585037e-3}(\pgfmathprintnumber [fixed,fixed zerofill,precision=0]{7.415181e1 })&\pgfmathprintnumber [fixed,fixed zerofill,precision=5]{5.7566803e-3}(\pgfmathprintnumber [fixed,fixed zerofill,precision=0]{1.0162979e2 })\\%
			\pgfutilensuremath {11}&\pgfmathprintnumber [fixed,fixed zerofill,precision=5]{4.9913605e-3}(\pgfmathprintnumber [fixed,fixed zerofill,precision=0]{5.4257446e1 })&\pgfmathprintnumber [fixed,fixed zerofill,precision=5]{4.9913605e-3}(\pgfmathprintnumber [fixed,fixed zerofill,precision=0]{5.4257446e1 })&\cellcolor {gray!50}\pgfmathprintnumber [fixed,fixed zerofill,precision=5]{4.6542297e-3}(\pgfmathprintnumber [fixed,fixed zerofill,precision=0]{8.1384125e1 })&\pgfmathprintnumber [fixed,fixed zerofill,precision=5]{4.1543076e-3}(\pgfmathprintnumber [fixed,fixed zerofill,precision=0]{1.154802e2 })\\%
			\pgfutilensuremath {12}&\pgfmathprintnumber [fixed,fixed zerofill,precision=5]{4.5206726e-3}(\pgfmathprintnumber [fixed,fixed zerofill,precision=0]{5.8941605e1 })&\pgfmathprintnumber [fixed,fixed zerofill,precision=5]{4.4266785e-3}(\pgfmathprintnumber [fixed,fixed zerofill,precision=0]{6.119609e1 })&\cellcolor {gray!50}\pgfmathprintnumber [fixed,fixed zerofill,precision=5]{4.0326645e-3}(\pgfmathprintnumber [fixed,fixed zerofill,precision=0]{9.4415695e1 })&\pgfmathprintnumber [fixed,fixed zerofill,precision=5]{3.3784439e-3}(\pgfmathprintnumber [fixed,fixed zerofill,precision=0]{1.3779922e2 })\\\bottomrule %
			\end {tabular}%

		\end{center}
		\caption{\label{TAB_numerical_results_Vsa_Vsb}
			Hybrid spin-dependent potentials $V^{sa}_{11}(r) / c_F$ and $V^{sb}_{10}(r) / c_F$ in units of the lattice spacing $a = 0.060 \, \text{fm}$ at flow radius $r_f = \sqrt{8 t_f} = 1.8 \, a$.
			The four columns correspond to the four fitting variants discussed in Section~\ref{sec:extraction_matrixelements}.
			Our main results shown in Figure~\ref{fig:potentials} were obtained by setting $T_\text{min} / a = 9$ and are shaded in gray.
			Potential values with separations $r \ltapprox 2 r_f + a = 4.6 \, a$ might contain sizable systematic errors due to overlapping gauge links and should be taken with caution.}
		\end{table}
		
		
		\begin{table}[p]
		\def\arraystretch{1.2}
		\begin{center}
			\begin {tabular}{c|l|l|l|l}%
			\toprule \text {method} & \, $T_\text {min}(t,r)$ & $T_\text {min} / a = 8$ & $T_\text {min} / a = 9$ & $T_\text {min} / a = 10$ \\ \hline & \multicolumn {4}{c}{} \vspace {-0.5cm} \\ $r/a$ & \multicolumn {4}{c}{$a^2 V^\text {mix}_{\Sigma _u^-}(r) m_Q / c_F$} \\ & \multicolumn {4}{c}{} \vspace {-0.5cm} \\\midrule %
			\pgfutilensuremath {1}&\pgfmathprintnumber [fixed,fixed zerofill,precision=5]{2.21073303e-2}(\pgfmathprintnumber [fixed,fixed zerofill,precision=0]{1.8449318e1 })&\pgfmathprintnumber [fixed,fixed zerofill,precision=5]{2.21073303e-2}(\pgfmathprintnumber [fixed,fixed zerofill,precision=0]{1.8449318e1 })&\cellcolor {gray!50}\pgfmathprintnumber [fixed,fixed zerofill,precision=5]{2.20782776e-2}(\pgfmathprintnumber [fixed,fixed zerofill,precision=0]{2.6957161e1 })&\pgfmathprintnumber [fixed,fixed zerofill,precision=5]{2.20269394e-2}(\pgfmathprintnumber [fixed,fixed zerofill,precision=0]{3.48993301e1 })\\%
			\pgfutilensuremath {2}&\pgfmathprintnumber [fixed,fixed zerofill,precision=5]{2.04143219e-2}(\pgfmathprintnumber [fixed,fixed zerofill,precision=0]{1.65356522e1 })&\pgfmathprintnumber [fixed,fixed zerofill,precision=5]{2.04143219e-2}(\pgfmathprintnumber [fixed,fixed zerofill,precision=0]{1.65356522e1 })&\cellcolor {gray!50}\pgfmathprintnumber [fixed,fixed zerofill,precision=5]{2.03294144e-2}(\pgfmathprintnumber [fixed,fixed zerofill,precision=0]{2.3570488e1 })&\pgfmathprintnumber [fixed,fixed zerofill,precision=5]{2.02362213e-2}(\pgfmathprintnumber [fixed,fixed zerofill,precision=0]{3.00254745e1 })\\%
			\pgfutilensuremath {3}&\pgfmathprintnumber [fixed,fixed zerofill,precision=5]{1.87151031e-2}(\pgfmathprintnumber [fixed,fixed zerofill,precision=0]{1.54052429e1 })&\pgfmathprintnumber [fixed,fixed zerofill,precision=5]{1.87151031e-2}(\pgfmathprintnumber [fixed,fixed zerofill,precision=0]{1.54052429e1 })&\cellcolor {gray!50}\pgfmathprintnumber [fixed,fixed zerofill,precision=5]{1.8608841e-2}(\pgfmathprintnumber [fixed,fixed zerofill,precision=0]{2.17283096e1 })&\pgfmathprintnumber [fixed,fixed zerofill,precision=5]{1.8499443e-2}(\pgfmathprintnumber [fixed,fixed zerofill,precision=0]{2.7150032e1 })\\%
			\pgfutilensuremath {4}&\pgfmathprintnumber [fixed,fixed zerofill,precision=5]{1.71987152e-2}(\pgfmathprintnumber [fixed,fixed zerofill,precision=0]{1.4403511e1 })&\pgfmathprintnumber [fixed,fixed zerofill,precision=5]{1.71987152e-2}(\pgfmathprintnumber [fixed,fixed zerofill,precision=0]{1.4403511e1 })&\cellcolor {gray!50}\pgfmathprintnumber [fixed,fixed zerofill,precision=5]{1.71002579e-2}(\pgfmathprintnumber [fixed,fixed zerofill,precision=0]{2.06500015e1 })&\pgfmathprintnumber [fixed,fixed zerofill,precision=5]{1.69958496e-2}(\pgfmathprintnumber [fixed,fixed zerofill,precision=0]{2.59806137e1 })\\\hline %
			\pgfutilensuremath {5}&\pgfmathprintnumber [fixed,fixed zerofill,precision=5]{1.58852081e-2}(\pgfmathprintnumber [fixed,fixed zerofill,precision=0]{1.4249527e1 })&\pgfmathprintnumber [fixed,fixed zerofill,precision=5]{1.58852081e-2}(\pgfmathprintnumber [fixed,fixed zerofill,precision=0]{1.4249527e1 })&\cellcolor {gray!50}\pgfmathprintnumber [fixed,fixed zerofill,precision=5]{1.57787933e-2}(\pgfmathprintnumber [fixed,fixed zerofill,precision=0]{2.05166321e1 })&\pgfmathprintnumber [fixed,fixed zerofill,precision=5]{1.56782684e-2}(\pgfmathprintnumber [fixed,fixed zerofill,precision=0]{2.59016495e1 })\\%
			\pgfutilensuremath {6}&\pgfmathprintnumber [fixed,fixed zerofill,precision=5]{1.46221008e-2}(\pgfmathprintnumber [fixed,fixed zerofill,precision=0]{1.46816788e1 })&\pgfmathprintnumber [fixed,fixed zerofill,precision=5]{1.46221008e-2}(\pgfmathprintnumber [fixed,fixed zerofill,precision=0]{1.46816788e1 })&\cellcolor {gray!50}\pgfmathprintnumber [fixed,fixed zerofill,precision=5]{1.45291367e-2}(\pgfmathprintnumber [fixed,fixed zerofill,precision=0]{2.09686508e1 })&\pgfmathprintnumber [fixed,fixed zerofill,precision=5]{1.44557419e-2}(\pgfmathprintnumber [fixed,fixed zerofill,precision=0]{2.63278427e1 })\\%
			\pgfutilensuremath {7}&\pgfmathprintnumber [fixed,fixed zerofill,precision=5]{1.34430771e-2}(\pgfmathprintnumber [fixed,fixed zerofill,precision=0]{1.62333984e1 })&\pgfmathprintnumber [fixed,fixed zerofill,precision=5]{1.34430771e-2}(\pgfmathprintnumber [fixed,fixed zerofill,precision=0]{1.62333984e1 })&\cellcolor {gray!50}\pgfmathprintnumber [fixed,fixed zerofill,precision=5]{1.33620758e-2}(\pgfmathprintnumber [fixed,fixed zerofill,precision=0]{2.36085587e1 })&\pgfmathprintnumber [fixed,fixed zerofill,precision=5]{1.33025131e-2}(\pgfmathprintnumber [fixed,fixed zerofill,precision=0]{2.96003036e1 })\\%
			\pgfutilensuremath {8}&\pgfmathprintnumber [fixed,fixed zerofill,precision=5]{1.24655914e-2}(\pgfmathprintnumber [fixed,fixed zerofill,precision=0]{1.67877579e1 })&\pgfmathprintnumber [fixed,fixed zerofill,precision=5]{1.24655914e-2}(\pgfmathprintnumber [fixed,fixed zerofill,precision=0]{1.67877579e1 })&\cellcolor {gray!50}\pgfmathprintnumber [fixed,fixed zerofill,precision=5]{1.2417984e-2}(\pgfmathprintnumber [fixed,fixed zerofill,precision=0]{2.58139954e1 })&\pgfmathprintnumber [fixed,fixed zerofill,precision=5]{1.23752518e-2}(\pgfmathprintnumber [fixed,fixed zerofill,precision=0]{3.39647675e1 })\\%
			\pgfutilensuremath {9}&\pgfmathprintnumber [fixed,fixed zerofill,precision=5]{1.13256378e-2}(\pgfmathprintnumber [fixed,fixed zerofill,precision=0]{1.90773773e1 })&\pgfmathprintnumber [fixed,fixed zerofill,precision=5]{1.13256378e-2}(\pgfmathprintnumber [fixed,fixed zerofill,precision=0]{1.90773773e1 })&\cellcolor {gray!50}\pgfmathprintnumber [fixed,fixed zerofill,precision=5]{1.13290863e-2}(\pgfmathprintnumber [fixed,fixed zerofill,precision=0]{3.10310135e1 })&\pgfmathprintnumber [fixed,fixed zerofill,precision=5]{1.13271484e-2}(\pgfmathprintnumber [fixed,fixed zerofill,precision=0]{4.27846909e1 })\\%
			\pgfutilensuremath {10}&\pgfmathprintnumber [fixed,fixed zerofill,precision=5]{1.05082703e-2}(\pgfmathprintnumber [fixed,fixed zerofill,precision=0]{2.09435501e1 })&\pgfmathprintnumber [fixed,fixed zerofill,precision=5]{1.05082703e-2}(\pgfmathprintnumber [fixed,fixed zerofill,precision=0]{2.09435501e1 })&\cellcolor {gray!50}\pgfmathprintnumber [fixed,fixed zerofill,precision=5]{1.0466835e-2}(\pgfmathprintnumber [fixed,fixed zerofill,precision=0]{3.35099716e1 })&\pgfmathprintnumber [fixed,fixed zerofill,precision=5]{1.04180984e-2}(\pgfmathprintnumber [fixed,fixed zerofill,precision=0]{4.53837662e1 })\\%
			\pgfutilensuremath {11}&\pgfmathprintnumber [fixed,fixed zerofill,precision=5]{9.3406982e-3}(\pgfmathprintnumber [fixed,fixed zerofill,precision=0]{2.18341827e1 })&\pgfmathprintnumber [fixed,fixed zerofill,precision=5]{9.3406982e-3}(\pgfmathprintnumber [fixed,fixed zerofill,precision=0]{2.18341827e1 })&\cellcolor {gray!50}\pgfmathprintnumber [fixed,fixed zerofill,precision=5]{9.2230301e-3}(\pgfmathprintnumber [fixed,fixed zerofill,precision=0]{3.39839783e1 })&\pgfmathprintnumber [fixed,fixed zerofill,precision=5]{9.0951614e-3}(\pgfmathprintnumber [fixed,fixed zerofill,precision=0]{4.44516373e1 })\\%
			\pgfutilensuremath {12}&\pgfmathprintnumber [fixed,fixed zerofill,precision=5]{8.5954056e-3}(\pgfmathprintnumber [fixed,fixed zerofill,precision=0]{2.38207321e1 })&\pgfmathprintnumber [fixed,fixed zerofill,precision=5]{8.5954056e-3}(\pgfmathprintnumber [fixed,fixed zerofill,precision=0]{2.38207321e1 })&\cellcolor {gray!50}\pgfmathprintnumber [fixed,fixed zerofill,precision=5]{8.3904343e-3}(\pgfmathprintnumber [fixed,fixed zerofill,precision=0]{3.74281921e1 })&\pgfmathprintnumber [fixed,fixed zerofill,precision=5]{8.2015381e-3}(\pgfmathprintnumber [fixed,fixed zerofill,precision=0]{4.97407379e1 })\\\bottomrule %
			\end {tabular}%
			
			\vspace{1.0cm}
			
			\begin {tabular}{c|l|l|l|l}%
			\toprule \text {method} & \, $T_\text {min}(t,r)$ & $T_\text {min} / a = 8$ & $T_\text {min} / a = 9$ & $T_\text {min} / a = 10$ \\ \hline & \multicolumn {4}{c}{} \vspace {-0.5cm} \\ $r/a$ & \multicolumn {4}{c}{$a^2 V^\text {mix}_{\Pi _u}(r) m_Q / c_F$} \\ & \multicolumn {4}{c}{} \vspace {-0.5cm} \\\midrule %
			\pgfutilensuremath {1}&\pgfmathprintnumber [fixed,fixed zerofill,precision=5]{2.28373032e-2}(\pgfmathprintnumber [fixed,fixed zerofill,precision=0]{1.76818619e1 })&\pgfmathprintnumber [fixed,fixed zerofill,precision=5]{2.28373032e-2}(\pgfmathprintnumber [fixed,fixed zerofill,precision=0]{1.76818619e1 })&\cellcolor {gray!50}\pgfmathprintnumber [fixed,fixed zerofill,precision=5]{2.28070984e-2}(\pgfmathprintnumber [fixed,fixed zerofill,precision=0]{2.59286575e1 })&\pgfmathprintnumber [fixed,fixed zerofill,precision=5]{2.2762352e-2}(\pgfmathprintnumber [fixed,fixed zerofill,precision=0]{3.3727745e1 })\\%
			\pgfutilensuremath {2}&\pgfmathprintnumber [fixed,fixed zerofill,precision=5]{2.09210663e-2}(\pgfmathprintnumber [fixed,fixed zerofill,precision=0]{1.50113373e1 })&\pgfmathprintnumber [fixed,fixed zerofill,precision=5]{2.09210663e-2}(\pgfmathprintnumber [fixed,fixed zerofill,precision=0]{1.50113373e1 })&\cellcolor {gray!50}\pgfmathprintnumber [fixed,fixed zerofill,precision=5]{2.08714981e-2}(\pgfmathprintnumber [fixed,fixed zerofill,precision=0]{2.20988998e1 })&\pgfmathprintnumber [fixed,fixed zerofill,precision=5]{2.08185272e-2}(\pgfmathprintnumber [fixed,fixed zerofill,precision=0]{2.85694122e1 })\\%
			\pgfutilensuremath {3}&\pgfmathprintnumber [fixed,fixed zerofill,precision=5]{1.85199509e-2}(\pgfmathprintnumber [fixed,fixed zerofill,precision=0]{1.26455002e1 })&\pgfmathprintnumber [fixed,fixed zerofill,precision=5]{1.85199509e-2}(\pgfmathprintnumber [fixed,fixed zerofill,precision=0]{1.26455002e1 })&\cellcolor {gray!50}\pgfmathprintnumber [fixed,fixed zerofill,precision=5]{1.84441376e-2}(\pgfmathprintnumber [fixed,fixed zerofill,precision=0]{1.85823593e1 })&\pgfmathprintnumber [fixed,fixed zerofill,precision=5]{1.83757782e-2}(\pgfmathprintnumber [fixed,fixed zerofill,precision=0]{2.3808693e1 })\\%
			\pgfutilensuremath {4}&\pgfmathprintnumber [fixed,fixed zerofill,precision=5]{1.61492386e-2}(\pgfmathprintnumber [fixed,fixed zerofill,precision=0]{1.08523026e1 })&\pgfmathprintnumber [fixed,fixed zerofill,precision=5]{1.61492386e-2}(\pgfmathprintnumber [fixed,fixed zerofill,precision=0]{1.08523026e1 })&\cellcolor {gray!50}\pgfmathprintnumber [fixed,fixed zerofill,precision=5]{1.60611801e-2}(\pgfmathprintnumber [fixed,fixed zerofill,precision=0]{1.59512405e1 })&\pgfmathprintnumber [fixed,fixed zerofill,precision=5]{1.59867706e-2}(\pgfmathprintnumber [fixed,fixed zerofill,precision=0]{2.04638062e1 })\\\hline %
			\pgfutilensuremath {5}&\pgfmathprintnumber [fixed,fixed zerofill,precision=5]{1.40251236e-2}(\pgfmathprintnumber [fixed,fixed zerofill,precision=0]{9.2754288e0 })&\pgfmathprintnumber [fixed,fixed zerofill,precision=5]{1.40251236e-2}(\pgfmathprintnumber [fixed,fixed zerofill,precision=0]{9.2754288e0 })&\cellcolor {gray!50}\pgfmathprintnumber [fixed,fixed zerofill,precision=5]{1.3937828e-2}(\pgfmathprintnumber [fixed,fixed zerofill,precision=0]{1.3870781e1 })&\pgfmathprintnumber [fixed,fixed zerofill,precision=5]{1.38627243e-2}(\pgfmathprintnumber [fixed,fixed zerofill,precision=0]{1.8262291e1 })\\%
			\pgfutilensuremath {6}&\pgfmathprintnumber [fixed,fixed zerofill,precision=5]{1.22054901e-2}(\pgfmathprintnumber [fixed,fixed zerofill,precision=0]{7.299591e0 })&\pgfmathprintnumber [fixed,fixed zerofill,precision=5]{1.22054901e-2}(\pgfmathprintnumber [fixed,fixed zerofill,precision=0]{7.299591e0 })&\cellcolor {gray!50}\pgfmathprintnumber [fixed,fixed zerofill,precision=5]{1.21289902e-2}(\pgfmathprintnumber [fixed,fixed zerofill,precision=0]{1.09861755e1 })&\pgfmathprintnumber [fixed,fixed zerofill,precision=5]{1.20555725e-2}(\pgfmathprintnumber [fixed,fixed zerofill,precision=0]{1.50429077e1 })\\%
			\pgfutilensuremath {7}&\pgfmathprintnumber [fixed,fixed zerofill,precision=5]{1.06239624e-2}(\pgfmathprintnumber [fixed,fixed zerofill,precision=0]{6.7854767e0 })&\pgfmathprintnumber [fixed,fixed zerofill,precision=5]{1.06239624e-2}(\pgfmathprintnumber [fixed,fixed zerofill,precision=0]{6.7854767e0 })&\cellcolor {gray!50}\pgfmathprintnumber [fixed,fixed zerofill,precision=5]{1.0535263e-2}(\pgfmathprintnumber [fixed,fixed zerofill,precision=0]{9.6236649e0 })&\pgfmathprintnumber [fixed,fixed zerofill,precision=5]{1.04497452e-2}(\pgfmathprintnumber [fixed,fixed zerofill,precision=0]{1.27036285e1 })\\%
			\pgfutilensuremath {8}&\pgfmathprintnumber [fixed,fixed zerofill,precision=5]{9.276947e-3}(\pgfmathprintnumber [fixed,fixed zerofill,precision=0]{7.4527893e0 })&\pgfmathprintnumber [fixed,fixed zerofill,precision=5]{9.276947e-3}(\pgfmathprintnumber [fixed,fixed zerofill,precision=0]{7.4527893e0 })&\cellcolor {gray!50}\pgfmathprintnumber [fixed,fixed zerofill,precision=5]{9.1528244e-3}(\pgfmathprintnumber [fixed,fixed zerofill,precision=0]{1.06965714e1 })&\pgfmathprintnumber [fixed,fixed zerofill,precision=5]{9.0492935e-3}(\pgfmathprintnumber [fixed,fixed zerofill,precision=0]{1.36527176e1 })\\%
			\pgfutilensuremath {9}&\pgfmathprintnumber [fixed,fixed zerofill,precision=5]{7.9690628e-3}(\pgfmathprintnumber [fixed,fixed zerofill,precision=0]{1.22302551e1 })&\pgfmathprintnumber [fixed,fixed zerofill,precision=5]{8.1256332e-3}(\pgfmathprintnumber [fixed,fixed zerofill,precision=0]{8.6112976e0 })&\cellcolor {gray!50}\pgfmathprintnumber [fixed,fixed zerofill,precision=5]{7.9690628e-3}(\pgfmathprintnumber [fixed,fixed zerofill,precision=0]{1.22302551e1 })&\pgfmathprintnumber [fixed,fixed zerofill,precision=5]{7.8580475e-3}(\pgfmathprintnumber [fixed,fixed zerofill,precision=0]{1.49221725e1 })\\%
			\pgfutilensuremath {10}&\pgfmathprintnumber [fixed,fixed zerofill,precision=5]{6.9883804e-3}(\pgfmathprintnumber [fixed,fixed zerofill,precision=0]{1.3011322e1 })&\pgfmathprintnumber [fixed,fixed zerofill,precision=5]{7.163742e-3}(\pgfmathprintnumber [fixed,fixed zerofill,precision=0]{9.206253e0 })&\cellcolor {gray!50}\pgfmathprintnumber [fixed,fixed zerofill,precision=5]{6.9883804e-3}(\pgfmathprintnumber [fixed,fixed zerofill,precision=0]{1.3011322e1 })&\pgfmathprintnumber [fixed,fixed zerofill,precision=5]{6.8667755e-3}(\pgfmathprintnumber [fixed,fixed zerofill,precision=0]{1.5979538e1 })\\%
			\pgfutilensuremath {11}&\pgfmathprintnumber [fixed,fixed zerofill,precision=5]{6.3380203e-3}(\pgfmathprintnumber [fixed,fixed zerofill,precision=0]{1.35471268e1 })&\pgfmathprintnumber [fixed,fixed zerofill,precision=5]{6.4792328e-3}(\pgfmathprintnumber [fixed,fixed zerofill,precision=0]{9.4564285e0 })&\cellcolor {gray!50}\pgfmathprintnumber [fixed,fixed zerofill,precision=5]{6.3380203e-3}(\pgfmathprintnumber [fixed,fixed zerofill,precision=0]{1.35471268e1 })&\pgfmathprintnumber [fixed,fixed zerofill,precision=5]{6.2420044e-3}(\pgfmathprintnumber [fixed,fixed zerofill,precision=0]{1.71462326e1 })\\%
			\pgfutilensuremath {12}&\pgfmathprintnumber [fixed,fixed zerofill,precision=5]{5.9100494e-3}(\pgfmathprintnumber [fixed,fixed zerofill,precision=0]{1.02195206e1 })&\pgfmathprintnumber [fixed,fixed zerofill,precision=5]{5.9100494e-3}(\pgfmathprintnumber [fixed,fixed zerofill,precision=0]{1.02195206e1 })&\cellcolor {gray!50}\pgfmathprintnumber [fixed,fixed zerofill,precision=5]{5.7650986e-3}(\pgfmathprintnumber [fixed,fixed zerofill,precision=0]{1.4647873e1 })&\pgfmathprintnumber [fixed,fixed zerofill,precision=5]{5.6446152e-3}(\pgfmathprintnumber [fixed,fixed zerofill,precision=0]{1.88308334e1 })\\\bottomrule %
			\end {tabular}%
		
		\end{center}
		\caption{\label{TAB_numerical_results_Vmix}
			Hybrid-quarkonium mixing potentials $V^\text{mix}_{\Sigma_u^-}(r) m_Q / c_F$ and $V^\text{mix}_{\Pi_u}(r) m_Q / c_F$ in units of the lattice spacing $a = 0.060 \, \text{fm}$ at flow radius $r_f = \sqrt{8 t_f} = 1.8 \, a$. 
			The four columns correspond to the four fitting variants discussed in Section~\ref{sec:extraction_matrixelements}.
			Our main results shown in Figure~\ref{fig:potentials} were obtained by setting $T_\text{min} / a = 9$ and are shaded in gray.
			Potential values with separations $r \ltapprox 2 r_f + a = 4.6 \, a$ might contain sizable systematic errors due to overlapping gauge links and should be taken with caution.}
		\end{table}
		
\begin{figure}[htb]
			\includegraphics[width=0.5\linewidth,page=2]{Figure2_tmin_eq_max_trf_9}
			\includegraphics[width=0.5\linewidth,page=1]{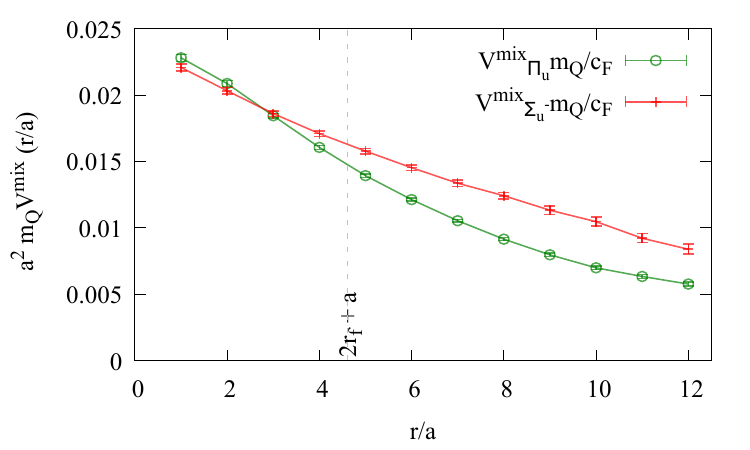}
			\caption{Hybrid spin-dependent potentials $V^{sa}_{11}(r) / c_F$ and $V^{sb}_{10}(r) / c_F$ (left) and hybrid-quarkonium mixing potentials $V^\text{mix}_{\Sigma_u^-}(r) m_Q / c_F$ and $V^\text{mix}_{\Pi_u}(r) m_Q / c_F$  (right) in units of the lattice spacing $a = 0.060 \, \text{fm}$ at flow radius $r_f = \sqrt{8 t_f} = 1.8 \, a$.}
			\label{fig:potentials}
		\end{figure}

		The computed potentials $V^{sa}_{11}(r) / c_F$, $V^{sb}_{10}(r) / c_F$, $V^\text{mix}_{\Sigma_u^-}(r) m_Q / c_F$ and $V^\text{mix}_{\Pi_u}(r) m_Q / c_F$ for $T_\text{min} / a = 9$ are shown in Figure~\ref{fig:potentials}.
		The matching coefficient $c_F(t_f,\mu)$, which translates from the gradient flow scheme at flow time $t_f$ to the $\overline{\text{MS}}$ scheme at scale $\mu$, is known up to one-loop order and can be found in Refs.\ \cite{Brambilla:2023vwm,Altenkort:2024spl}.
		We note again that the parallel transporters in temporal direction including the chromomagnetic field insertions, which appear in the generalized Wilson loops, are smeared due to the application of gradient flow.
		For separations $r \ltapprox 2 r_f + a = 4.6 \, a$, indicated by vertical dashed lines, discretization errors may become sizable due to overlapping gauge links from opposite parallel transporters. 
		Results in this region should be taken with caution (see also Section~\ref{sec:gradientflow}).
		
		Even though the results presented in Table~\ref{TAB_numerical_results_Vsa_Vsb}, Table~\ref{TAB_numerical_results_Vmix} and Figure~\ref{fig:potentials} were generated at fixed finite lattice spacing $a = 0.060 \, \text{fm}$ and flow radius $r_f = \sqrt{8 t_f} = 1.8 \, a$, we expect that they are quite similar to continuum and zero-flow time extrapolated results.
		The matching coefficient $c_F(t_f.\mu)$, which can be calculated according to Eq.\ (4.68) of Ref.\ \cite{Brambilla:2023vwm}, is close to $1$ for suitably chosen scale $\mu \approx 1 / \sqrt{2 t_f e^{\gamma_E}}$ and lattice discretization errors should be small for $r \gtapprox 2 r_f + a$ as discussed in Section~\ref{sec:gradientflow}.
		Since the potentials $V^{sa}_{11}(r)$, $V^{sb}_{10}(r)$, $V^\text{mix}_{\Sigma_u^-}(r)$ and $V^\text{mix}_{\Pi_u}(r)$ were previously modeled by combining predictions from pNRQCD and QCD effective string theory for their small and large $r$ behavior, there were open questions even on a qualitative level concerning their $r$-dependence (for example it was unclear, whether the mixing potentials $V^\text{mix}_{\Sigma_u^-}(r)$ and $V^\text{mix}_{\Pi_u}(r)$ have a positive or negative sign, see Ref.~\cite{Oncala:2017hop}).
		We, thus, believe that our lattice gauge theory results, even though not continuum and zero-flow time extrapolated, provide important new information, which could e.g.\ motivate refined computations of heavy hybrid meson spectra.
		Our results also indicate that a more rigorous extraction within our lattice setup is feasible.
		This will, however, require results for several lattice spacings and flow times and a corresponding combined continuum and zero-flow time extrapolation, which is beyond the scope of this paper, but planned as a follow-up project.

		
		\subsection{Alternative analysis method}
		
		One can think of other strategies besides the ratio defined in Eq.\ \eqref{eq:ratioRBk} to extract the matrix elements appearing in Eqs.\ \eqref{eq:Vsa} to \eqref{eq:VmixSigma}.
		One such alternative is to define
		\begin{equation}\label{eq:alternativeR}
			R^{B_k}_{2;\Lambda_\eta^\epsilon{\Lambda_\eta^\epsilon}'} (t;r,T) = \frac{W^{B_k}_{\Lambda_\eta^\epsilon{\Lambda_\eta^\epsilon}'}(t;r,T)}{{W_{\Lambda_\eta^\epsilon}(r,T/2-t)W_{ {\Lambda_\eta^\epsilon}'}(r,T/2+t)}} .
		\end{equation}
		From the spectral decompositions (\ref{EQN_specWB}) and (\ref{EQN_specW}) one can derive
		\begin{equation}
			\lim_{T\to \infty} R^{B_k}_{2;\Lambda_\eta^\epsilon{\Lambda_\eta^\epsilon}'} (t;r,T) = \frac{
			\bra{0,\Lambda_\eta^\epsilon} B^\text{lattice}_k \ket{0,{\Lambda_\eta^\epsilon}'}(r)}
			{\bra{0,{{\Lambda_\eta^\epsilon}}} O_{{\Lambda_\eta^\epsilon}}^\text{lattice} | \Omega \rangle
\langle \Omega | (O_{{\Lambda_\eta^\epsilon}^\prime}^\text{lattice})^\dagger\ket{0,{{\Lambda_\eta^\epsilon}^\prime}}}
			,
		\end{equation}
		as well as
		\begin{equation}
			\label{EQN037} \lim_{T\to \infty} W_{{\Lambda_\eta^\epsilon}}(r,T) = 
			\Big(\langle 0,{{\Lambda_\eta^\epsilon}} | O_{{\Lambda_\eta^\epsilon}}^\text{lattice} | \Omega \rangle\Big)^2 e^{-V_{{0,{\Lambda_\eta^\epsilon}}}(r)T} = 
			\Big(\langle \Omega | (O_{{\Lambda_\eta^\epsilon}}^\text{lattice})^\dagger | 0,{{\Lambda_\eta^\epsilon}} \rangle\Big)^2	e^{-V_{{0,{\Lambda_\eta^\epsilon}}}(r)T} ,
		\end{equation}
		where the ground state overlaps $\langle 0,{{\Lambda_\eta^\epsilon}} | O_{{\Lambda_\eta^\epsilon}}^\text{lattice} | \Omega \rangle = \langle \Omega | (O_{{\Lambda_\eta^\epsilon}}^\text{lattice})^\dagger | 0,{{\Lambda_\eta^\epsilon}} \rangle$ are positive, because of Eq.\ (\ref{eq:groundstate}).

		To cross-check our results from Section~\ref{sec:extraction_matrixelements}, we used these equations as follows.
		We first fitted exponential functions to lattice results for
		the Wilson loops $W_{\Lambda_\eta^\epsilon}(r,T)$ and $W_{{\Lambda_\eta^\epsilon}'}(r,T)$, to determine the ground state overlaps
		$\bra{0,{{\Lambda_\eta^\epsilon}}} O_{{\Lambda_\eta^\epsilon}}^\text{lattice} | \Omega \rangle$ and $\langle \Omega | (O_{{\Lambda_\eta^\epsilon}^\prime}^\text{lattice})^\dagger\ket{0,{{\Lambda_\eta^\epsilon}^\prime}}$ according to Eq.\ (\ref{EQN037}).
		A matrix element can then be obtained from a simple fit of a constant to lattice results for the ratio \eqref{eq:alternativeR} in the plateau region at large $T$ and a subsequent division by the previously determined ground state overlaps.
		The results and their statistical errors are comparable to those obtained by our main analysis based on 
		ratio \eqref{eq:ratioRBk} and presented in Section~\ref{sec:extraction_matrixelements}.
		This is reassuring, but not surprising, because we use the same lattice data and perform a full jackknife analysis for each of the two analysis methods.

		
		\subsection{Flow time dependence of correlator ratios and matrix elements}\label{sec:gradientflowtime_dependence}

\begin{figure}[htb]
			\centering
			\includegraphics[width=0.70\linewidth]{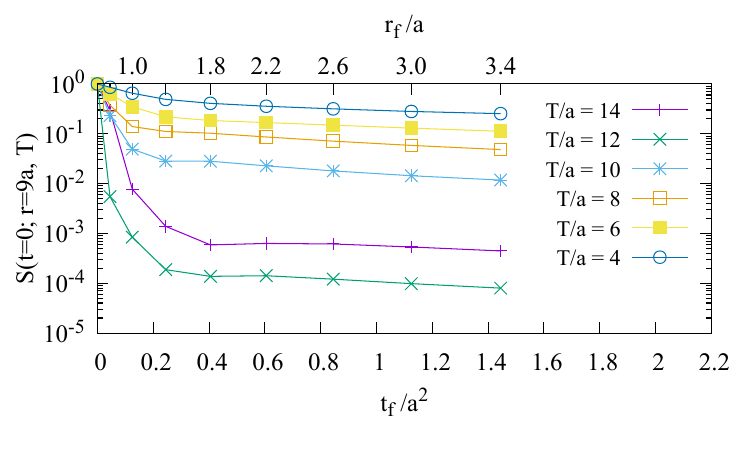}
			\caption{Relative error $S(t=0;r=9a,T)$ as defined in Eq.\ (\ref{EQN_S}) as function of the flow time $t_f / a^2$ for several $T$.}
			\label{fig:signaltonoise_RBk}
		\end{figure}

		By using gradient flow one can improve the signal quality of the correlator ratios $R^{B_k}_{\Lambda_\eta^\epsilon{\Lambda_\eta^\epsilon}'}(t;r,T)$ significantly.
		To demonstrate this, we plot in Figure~\ref{fig:signaltonoise_RBk}
		\begin{equation}
			\label{EQN_S} S(t;r,T) = \frac{\Delta R^{B_x}_{\Sigma_g^+ \Pi_u^+}(t;r,T)\Big|_{t_f}}{\Delta R^{B_x}_{\Sigma_g^+ \Pi_u^+}(t;r,T)\Big|_{t_f = 0}} ,
		\end{equation}
		the statistical error of $R^{B_x}_{\Sigma_g^+ \Pi_u^+}(t;r,T)$ at flow time $t_f \geq 0$ in units of the statistical error at flow time $0$ as function of $t_f$. We show curves for several $T$ with $t = 0$ and $r/a = 9$.
		It is obvious that gradient flow reduces statistical errors drastically. In particular for large $T$ there can be a gain of a few orders of magnitude. To maximally exploit this error reduction, one should use a flow time $t_f / a^2 \gtapprox 0.4$ or equivalently a flow radius $r_f / a \gtapprox 1.8$. On the other hand, a large flow radius has the drawback of possibly introducing sizable systematic errors as discussed in Section~\ref{sec:gradientflow}, if $T$ and $r$ are not sufficiently large. The former is a less severe problem, since the potentials are extracted from ratios at large $T$ (see Eq.\ (\ref{EQN_R_matrix_element})). However, the restriction on $r$ discussed in Section~\ref{sec:gradientflow}, $r \gtapprox 2 r_f + a$, is not ideal, because one is typically interested in both small and large separations $r$. In Section~\ref{sec:extraction_matrixelements}, where we presented the main results for the potentials $V^{sa}_{11}(r) / c_F$, $V^{sb}_{10}(r) / c_F$, $V^\text{mix}_{\Sigma_u^-}(r) m_Q / c_F$ and $V^\text{mix}_{\Pi_u}(r) m_Q / c_F$, we chose $r_f / a = 1.8$ as compromise.

\begin{figure}[htb]
			\begin{center}
			\includegraphics[width=0.49\linewidth,page=3]{Figure4}
			\includegraphics[width=0.49\linewidth,page=4]{Figure4}
			\includegraphics[width=0.49\linewidth,page=1]{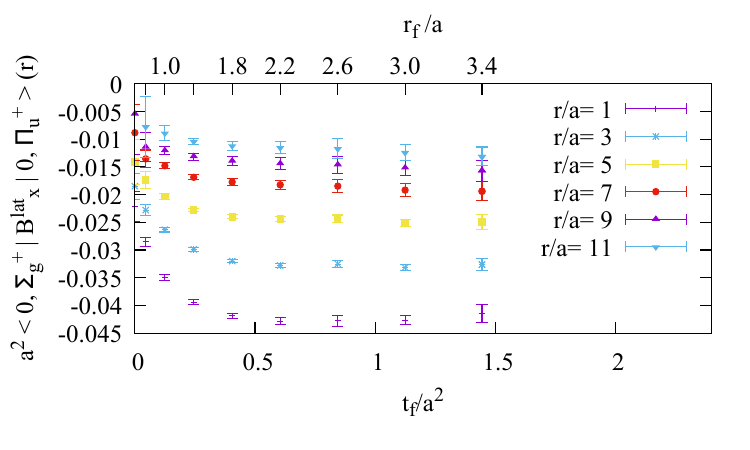}
			\includegraphics[width=0.49\linewidth,page=2]{Figure4}
			\end{center}
			\caption{The matrix elements $a^2 \bra{0 , \Pi_u^-} B^\text{lattice}_z \ket{0 , \Pi_u^+}(r)$ (top left), $a^2 \bra{0 , \Sigma_u^-} B^\text{lattice}_y \ket{0 , \Pi_u^+}(r)$ (top right), $a^2 \bra{0 , \Sigma_g^+} B^\text{lattice}_x \ket{0 , \Pi_u^+}(r)$ (bottom left) and $a^2 \bra{0 , \Sigma_g^+} B^\text{lattice}_z \ket{0 , \Sigma_u^-}(r)$ (bottom right) for selected separations $r$ as functions of the flow time $t_f / a^2$.}
			\label{fig:flowtimedependence_potentials_fixedr}
		\end{figure}

		In Figure~\ref{fig:flowtimedependence_potentials_fixedr} we show the matrix elements $\bra{0 , \Pi_u^-} B^\text{lattice}_z \ket{0 , \Pi_u^+}(r)$, $\bra{0 , \Sigma_u^-} B^\text{lattice}_y \ket{0 , \Pi_u^+}(r)$, \\ $\bra{0 , \Sigma_g^+} B^\text{lattice}_x \ket{0 , \Pi_u^+}(r)$ and $\bra{0 , \Sigma_g^+} B^\text{lattice}_z \ket{0 , \Sigma_u^-}(r)$ for selected separations $r$ as functions of the flow time $t_f$ (the matrix elements were extracted with $T_\text{min} / a = 9$ as discussed in Section~\ref{sec:extraction_matrixelements}).
		For flow time $t_f / a^2 \gtapprox 0.4$ their dependence on $t_f$ is very weak, i.e.\ within statistical errors the data points exhibit an almost constant behavior for each of the four matrix elements. 

		To relate these matrix elements to the potentials $V^{sa}_{11}(r)$, $V^{sb}_{10}(r)$, $V^\text{mix}_{\Sigma_u^-}(r)$ and $V^\text{mix}_{\Pi_u}(r)$ one has to divide by the matching coefficient $c_F(t_f,\mu)$ (see Eqs.\ (\ref{eq:Vsa}) to (\ref{eq:VmixSigma})).
		Assuming that discretization errors are small, the logarithmic flow time dependence of $c_F(t_f,\mu)$ in perturbation theory should cancel the observed rather weak flow time dependence of the lattice results for the matrix elements at the considered order of perturbation theory \footnote{The matching coefficient $c_F(t_f,\mu)$ from the gradient flow scheme at flow time $t_f$ to the $\overline{\text{MS}}$ scheme at scale $\mu$ is known up to one-loop order and can be found in Refs.\ \cite{Brambilla:2023vwm,Altenkort:2024spl}.}.
		It is, thus, expected that a combined continuum and zero-flow time extrapolation of $c_f(t_f,\mu) \bra{0,\Lambda_\eta^\epsilon} B_k \ket{0,{\Lambda_\eta^\epsilon}'}(r)$, as outlined in Section~\ref{sec:gradientflow}, can be carried out in a controlled way and will yield stable finite results for the hybrid spin-dependent potentials and hybrid-quarkonium mixing potentials.
		In the near future we plan to generate lattice data for the matrix elements for several lattice spacings and flow times and to perform such an $a \rightarrow 0$ and $t_f \rightarrow 0$ extrapolation.

	
	\clearpage
	
	
	\clearpage
	\section{Summary and conclusions}\label{sec:conclusion}
	In this work we provide the first lattice gauge theory results for the $(1 /m_Q)^1$ hybrid spin-dependent potentials $V^{sa}(r) / c_F$ and $V^{sb}(r) / c_F$ as well as for the hybrid-quarkonium mixing potentials $V^\text{mix}_{\Pi_u}(r) m_Q / c_F$ and $V^\text{mix}_{\Sigma_u^-}(r) m_Q / c_F$.
	The main results are shown in Figure~\ref{fig:potentials} and collected in Table~\ref{TAB_numerical_results_Vsa_Vsb} and Table~\ref{TAB_numerical_results_Vmix} in the columns highlighted in gray.
	The results correspond to gauge group SU(3), lattice spacing $a = 0.060 \,\text{fm}$ and gradient flow radius $r_f / a = 1.8$.
	We chose this value for $r_f$, to achieve, on the one hand, a strong suppression of statistical fluctuations and, on the other hand, to avoid sizable discretization errors for $r/a \gtapprox 5$ due to overlapping gauge links from opposite temporal lines of the computed generalized Wilson loops.
	
	Reliable parametrizations of $(1/m_Q)^1$ hybrid spin-dependent potentials and hybrid-quarkonium mixing potentials are essential for precise predictions of heavy hybrid meson spectra within BOEFT (see e.g.\ Refs.\ \cite{Oncala:2017hop,Brambilla:2018pyn,Brambilla:2019jfi,Soto:2023lbh}).
	The lattice data points we provide in this work constrain such parameterizations in the region $0.30 \, \text{fm} \ltapprox r \ltapprox 0.72 \, \text{fm}$ and, thus, offer valuable insights into the behavior of these potentials.
	A comparison of our results from Figure~\ref{fig:potentials} with parametrizations used in the literature \cite{Soto:2017one,Oncala:2017hop,Brambilla:2018pyn,Brambilla:2019jfi,Soto:2023lbh} reveals discrepancies.
	For example, Ref.~\cite{Oncala:2017hop} explored mixing effects by considering several different choices for some of the parameters which are either unknown or have large uncertainties or unknown signs.
	However, none of these variants is reasonably similar to our lattice results. 
	We also fitted an ansatz to our lattice data for the hybrid spin-dependent potentials and hybrid-quarkonium mixing potentials, which is inspired by the parametrizations used in Refs. \cite{Oncala:2017hop,Soto:2023lbh} and incorporates both predictions from pNRQCD for small separations and expectations from QCD effective string theory for large separations.
	While the functional forms of these fit functions are suited to parametrize our data, the extracted parameter values deviate significantly from those used in Refs.\ \cite{Soto:2017one,Oncala:2017hop,Brambilla:2018pyn,Brambilla:2019jfi,Soto:2023lbh}.
	These findings indicate that existing determinations of spin and mixing effects in heavy hybrid meson spectra based on these parametrizations should be revisited and they emphasize the necessity of first-principles lattice gauge theory results to rigorously constrain and refine these parametrizations, which will be the focus of upcoming future work.
	
	We note again that the results presented in this work correspond to a single lattice spacing $a = 0.060 \,\text{fm}$ and gradient flow radius $r_f / a = 1.8$.
	Even though we expect that these results are already close to continuum results at zero flow radius (see the discussion at the end of Section~\ref{sec:extraction_matrixelements}), such a combined $a \to 0$ and $r_f \to 0$ extrapolation of the hybrid spin-dependent potentials $V^{sa}(r)$ and $V^{sb}(r)$ and of the hybrid-quarkonium mixing potentials $V^\text{mix}_{\Pi_u}(r) m_Q$ and $V^\text{mix}_{\Sigma_u^-}(r) m_Q$ will be mandatory for rigorous renormalized continuum results.
	We plan to continue this project by performing analogous lattice gauge theory computations for several values of $a$ and $r_f$ and carry out a corresponding combined extrapolation.


\clearpage

\section*{Acknowledgements}

	We acknowledge useful discussions with Nora Brambilla, Michael Eichberg, Abhishek Mohapatra, Joan Soto and Antonio Vairo.
	
	M.W.\ acknowledges support by the Deutsche Forschungsgemeinschaft (DFG, German Research Foundation) -- project number 550503820.
	M.W.\ acknowledges support by the Heisenberg Programme of the Deutsche Forschungsgemeinschaft (DFG, German Research Foundation) -- project number 399217702.
	
	Calculations on the GOETHE-NHR and on the FUCHS-CSC  high-performance computers of the Frankfurt University were conducted for this research. We would like to thank HPC-Hessen, funded by the State Ministry of Higher Education, Research and the Arts, for programming advice.


	\clearpage
	
	\bibliographystyle{utphys.bst}
	\providecommand{\href}[2]{#2}\begingroup\raggedright\endgroup

\end{document}